\documentclass{iopjournal}

\begin{document}

\articletype{Paper} 

\title{Tensor network investigation of the monomer-dimer model on the honeycomb lattice}

\author{De-Zhang Li$^{1,*}$\orcid{0000-0001-8039-449X}, Jie Liu$^{2,*}$\orcid{0000-0002-1436-8901} and Xin Wang$^{2,3,\dag}$\orcid{0000-0003-2971-5088}}

\affil{$^1$Quantum Science Center of Guangdong-Hong Kong-Macao Greater Bay Area, Shenzhen 518045, China}

\affil{$^2$Department of Physics, City University of Hong Kong, Hong Kong SAR, China}

\affil{$^3$City University of Hong Kong Shenzhen Research Institute, Shenzhen 518057, China}

\affil{$^*$These authors contributed equally to the paper.}

\affil{$^\dag$Author to whom any correspondence should be addressed.}

\email{x.wang@cityu.edu.hk}


\begin{abstract}
The monomer-dimer model is one of the most well-known unsolved lattice models. In this paper we study the monomer-dimer model on the honeycomb lattice using the tensor network method, in the case that the dimer and monomer activities are 1. The monomer-dimer configurations are exactly mapped into the ground states of the antiferromagnetic Ising model on the Kagom\'e lattice in the critical field $H_{\rm{ex}}=4J$, and the tensor network is constructed based on the local ground states of each Ising triangle. The VUMPS approach is employed to contract the tensor network, providing a high-precision result of the monomer-dimer problem. We also revisit the edge coloring problem on the honeycomb lattice and discuss its relationship to the monomer-dimer model. Finally we formulate the monomer-dimer problem in the language of the sixteen-vertex model, and discuss the non-integrability of the general monomer-dimer model and the integrability of the pure dimer model.
\end{abstract}

\section{Introduction}   \label{intro}
The monomer-dimer model was introduced in the 1930s \cite{RN577, RN400}, and has since become one of the most well-known statistical lattice models. On a given lattice, the monomer-dimer problem can be formulated as counting the number of ways of placing dimers and monomers, where each dimer occupies a pair of nearest-neighbour sites and each monomer occupies a single site, so that each site is occupied exactly by either one dimer or one monomer. When the model contains only dimers, it is referred to as the pure dimer covering model. The monomer-dimer model is not only important in statistical physics, but also of significant interest in the field of combinatorics. 

Approximate methods were employed in the early study of the monomer-dimer model \cite{RN543, RN578, RN544, RN545}. In 1961, the exact solution of the pure dimer model on the square lattice was obtained by the Pfaffian method \cite{RN136, RN137, RN138}. Ever since, the pure dimer models on various two-dimensional lattices have been exactly solved using a variety of approaches such as the Pfaffian \cite{RN139, RN397, RN143, RN486, RN145, RN380, RN382, RN465, RN387, RN389}, the transfer matrix \cite{RN141}, mapping into the free-fermion model \cite{RN140, RN129, RN144, RN393}, and mapping into the Ising model on the dual lattice \cite{RN157, RN385, RN158}. However, the general monomer-dimer model remains unsolved, except for some special cases with certain restrictions for the positions or number of monomers \cite{RN401, RN402, RN476, RN580, RN406, RN356}. It has been demonstrated that the monomer-dimer model on a two-dimensional lattice is \#$P$-complete \cite{RN355}. The monomer-dimer model, together with other unsolved problems such as the three-dimensional cubic lattice Ising model, serves as an important benchmark problem in the statistical physics of lattice systems. Many theoretical and numerical studies of the monomer-dimer model have been conducted \cite{RN146, RN147, RN199, RN189, RN148, RN149, RN375, RN407, RN408, RN151, RN347, RN357, RN579, RN660}. The monomer-dimer model therefore provides a useful benchmark for theoretical and numerical methods.

In this paper we focus on the monomer-dimer model on the honeycomb lattice, in the case that the dimer and monomer activities are 1. A remarkable estimate of the solution was obtained via the Bethe approximation by Nagle \cite{RN147}, which was rederived by Isakov \textit{et al.} \cite{RN104}. Deep connections have been shown between the monomer-dimer model and other lattice models. As pointed out by Ref.~\cite{RN149}, the monomer-dimer model on a lattice $A$ is closely related to the hard-core lattice gas model on the line graph of $A$. For example, the monomer-dimer model on the honeycomb lattice is equivalent to the hard-core lattice gas on the Kagom\'e lattice \cite{RN150}. The relation of the monomer-dimer model on the honeycomb lattice with the Ising models on the honeycomb and Kagom\'e lattices has also been studied \cite{RN117, RN104, RN120, RN225}. It is also obvious that the monomer-dimer model can be mapped into the edge coloring problem. In this work we employ the method of mapping from the monomer-dimer configurations into the ground states of the antiferromagnetic Ising model on the Kagom\'e lattice in the critical field. That is, the solution of our problem is exactly the residual entropy of the antiferromagnetic Kagom\'e Ising model in the critical field.

Numerical results of this residual entropy from Wang-Landau sampling \cite{RN120}, Monte Carlo simulation \cite{RN567} and tensor network method \cite{RN354} have been reported. In this work we provide a new tensor network study of this problem. Tensor networks offer a powerful framework for tackling unsolved lattice models and are therefore well suited to the present calculation \cite{RN281, RN203, RN282}. Initially developed from the transfer matrix techniques \cite{RN199, RN200}, tensor network methods have found broad applications in statistical mechanics. The partition function of a lattice model can usually be expressed as the contraction of a tensor network. Since the residual entropy can be seen as the zero-temperature limit of the partition function, it is natural to apply the tensor network to evaluate the residual entropy \cite{RN103, RN181, RN354, RN733}. Obviously, it is convenient to construct a tensor network specifically for the ground states in the residual entropy problem. In this work we follow closely the tensor network representation for the ground states of the antiferromagnetic Kagom\'e Ising model proposed in Ref.~\cite{RN181}. 

The remainder of this paper is organized as follows. In Sec.~\ref{Ising} we introduce the mapping from the monomer-dimer model on the honeycomb lattice into the antiferromagnetic Ising model on the Kagom\'e lattice in the critical field. In Sec.~\ref{tensor} the tensor network representation for the ground states and the numerical results of contraction in the thermodynamic limit are presented. Our result is compared with the previous estimates. We further revisit the edge coloring problem on the honeycomb lattice and discuss the relation with other statistical models including the monomer-dimer model in Sec.~\ref{coloring}. Discussion and summary are given in Sec.~\ref{summary}.

\section{Mapping into the antiferromagnetic Kagom\'e Ising model}   \label{Ising}
\begin{figure}
 \centering
        \includegraphics[width=0.45\textwidth]{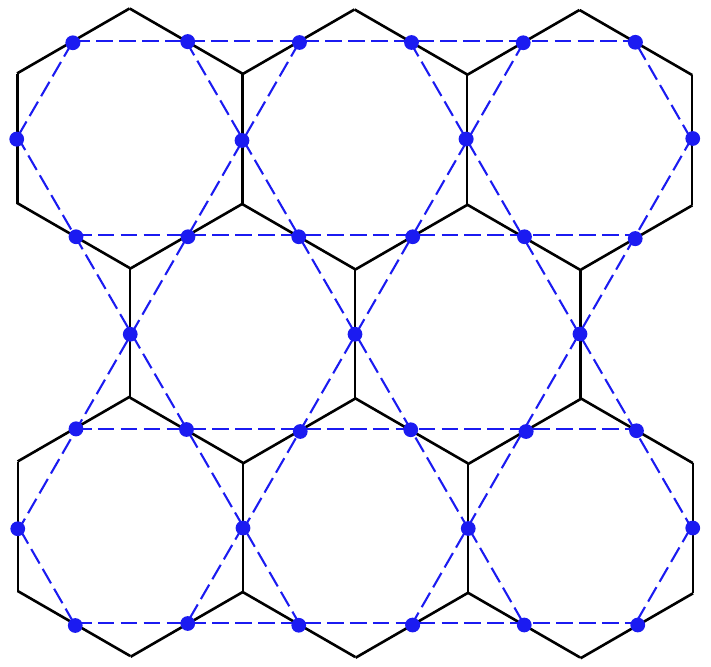} 
        \caption{Mapping from the monomer-dimer model on the honeycomb lattice into the Ising model on the Kagom\'e lattice. The honeycomb lattice is marked in solid black lines and the Kagom\'e lattice is marked in dashed blue lines.}  \label{fig1}
\end{figure}

The ground states of an antiferromagnetic Ising model in the critical field are equivalent to the configurations of the hard-core lattice gas on the same lattice \cite{RN581, RN582, RN471}. Since the monomer-dimer model can be transformed into the hard-core lattice gas model on the line graph \cite{RN149}, it can also be mapped into the corresponding antiferromagnetic Ising model in the critical field. Here we employ this method to map the monomer-dimer model on the honeycomb lattice into the antiferromagnetic Ising model on the Kagom\'e lattice in the critical field. By placing a spin on each edge of the honeycomb lattice and connecting the nearest-neighbour spins, we obtain a Kagom\'e lattice Ising model, as shown in Figure \ref{fig1}. The monomer-dimer configurations can be mapped into the Ising spin states, by assigning value $-1$ to each spin on the edge covered by a dimer and $+1$ to that on the uncovered edge.

The antiferromagnetic Ising models in the critical field on various lattices, including the square, honeycomb, triangular and simple cubic lattices, have been studied \cite{RN399, RN571, RN568, RN569}. Determining the critical point of an antiferromagnetic Ising model in a field is more complicated than that for the ferromagnetic model, as the well-known Lee-Yang circle theorem \cite{RN57} does not apply to the antiferromagnetic case. Fortunately, it can be proved from Theorems 2.1 and 3.1 of Ref.~\cite{RN570}, which investigate the Lee-Yang zeros of certain spin systems, that the antiferromagnetic Kagom\'e Ising model in a physical (real) field has no phase transition. Hence, there is no physical critical temperature for the antiferromagnetic Kagom\'e Ising model in the critical field. In the zero-temperature limit the system reduces to the ground states, which are shown below.

We denote the antiferromagnetic interaction by $J~(J>0)$ and the magnetic field by $H_{\rm{ex}}$. The Hamiltonian can be expressed as
\begin{equation}
H = \sum\limits_{\rm{tri}} J\left( s_1 s_2 + s_2 s_3 + s_1 s_3 \right) - H_{\rm{ex}}\sum\limits_i {s_i} = \sum\limits_{\rm{tri}} \left[ J\left( s_1 s_2 + s_2 s_3 + s_1 s_3 \right) - \frac{1}{2} H_{\rm{ex}}\left( s_1 + s_2 + s_3 \right) \right],    \label{eq1}
\end{equation}
where $\sum\nolimits_{\rm{tri}} {}$ is the summation over all triangles and $\left\{ s_1, s_2, s_3,~s_i = \pm 1 \right\}$ represent the three spins in each triangle. The Kagom\'e Ising model has been exactly solved in the cases of $H_{\rm{ex}}=0$ \cite{RN121, RN82} and $H_{\rm{ex}}=i(\pi/2)k_B T$ \cite{RN558, RN671}. Here we consider the case of a physical (real) field, for which no mathematically exact solution is currently known. We can set $H_{\rm{ex}}>0$, and the case that $H_{\rm{ex}}<0$ is symmetric.

The ground states of the Kagom\'e Ising model are determined by the local ground states of each triangle. For convenience we use the notation $2 \times \left(+1\right) + 1 \times \left(-1\right)$ to represent the configuration of a triangle that consists of two $+1$ spins and one $-1$ spin. The ground states in the presence of a field can be verified as follows: \\
\noindent (\romannumeral1). $H_{\rm{ex}}=0$, the local ground states are $2 \times \left(+1\right) + 1 \times \left(-1\right)$ and $2 \times \left(-1\right) + 1 \times \left(+1\right)$. The residual entropy was obtained by taking the zero-temperature limit of the exact solution \cite{RN82}. \\
\noindent (\romannumeral2). $0<H_{\rm{ex}}<4J$, only the configurations $2 \times \left(+1\right) + 1 \times \left(-1\right)$ remain as local ground states. Since each $-1$ spin represents a dimer and each $+1$ spin stands for an edge uncovered by a dimer, we can verify that this case is equivalent to the pure dimer model on the honeycomb lattice \cite{RN140}. The ground state degeneracy is reduced but still extensive.  \\
\noindent (\romannumeral3). $H_{\rm{ex}}=4J$, the local ground states in this critical field are $2 \times \left(+1\right) + 1 \times \left(-1\right)$ and $3 \times \left(+1\right)$. By regarding each triangle with configuration $3 \times \left(+1\right)$ as a site occupied by a monomer (none of the edges incident to this site is covered by a dimer), we can see that the ground states are exactly equivalent to the monomer-dimer configurations on the honeycomb lattice. \\
\noindent (\romannumeral4). $H_{\rm{ex}}>4J$, the local ground state is $3 \times \left(+1\right)$. The ground state of the system contains only one complete ordered configuration. 

We can see that the solution of the monomer-dimer model on the honeycomb lattice, with both dimer and monomer activities being 1, is transformed into the residual entropy of the antiferromagnetic Kagom\'e Ising model in the critical field $H_{\rm{ex}}=4J$. Let the number of sites on the honeycomb lattice be $N$, and the number of sites on the Kagom\'e lattice be $\frac{3}{2}N$. The solution of our problem can be expressed as $\lim_{N \to \infty} \frac{1}{N}\ln Z$, where $Z$ is the partition function of the monomer-dimer model and, equivalently, the ground state degeneracy of the antiferromagnetic Kagom\'e Ising model in the critical field. In the following section we present the tensor network construction specifically for the ground states in the critical field, and show our numerical results.

\section{Tensor network investigation}   \label{tensor}
The evaluation of residual entropy of a lattice model is a counting problem of the ground states. Hence, it is beneficial to design a tensor network representation specifically for the ground states, which is usually the zero-temperature limit of the standard construction \cite{RN576}. References \cite{RN181, RN345} have proposed a tensor network representation depending on the local rule for the Kagom\'e Ising ground states in the zero field, which was reformulated in Ref.~\cite{RN336}. Here we follow the idea of Refs.~\cite{RN181, RN345} and use a slightly different construction for the ground states in the critical field $H_{\rm{ex}}=4J$.

\subsection{Tensor network construction}   \label{constr}
\begin{figure}
 \centering
        \includegraphics[width=0.8\textwidth]{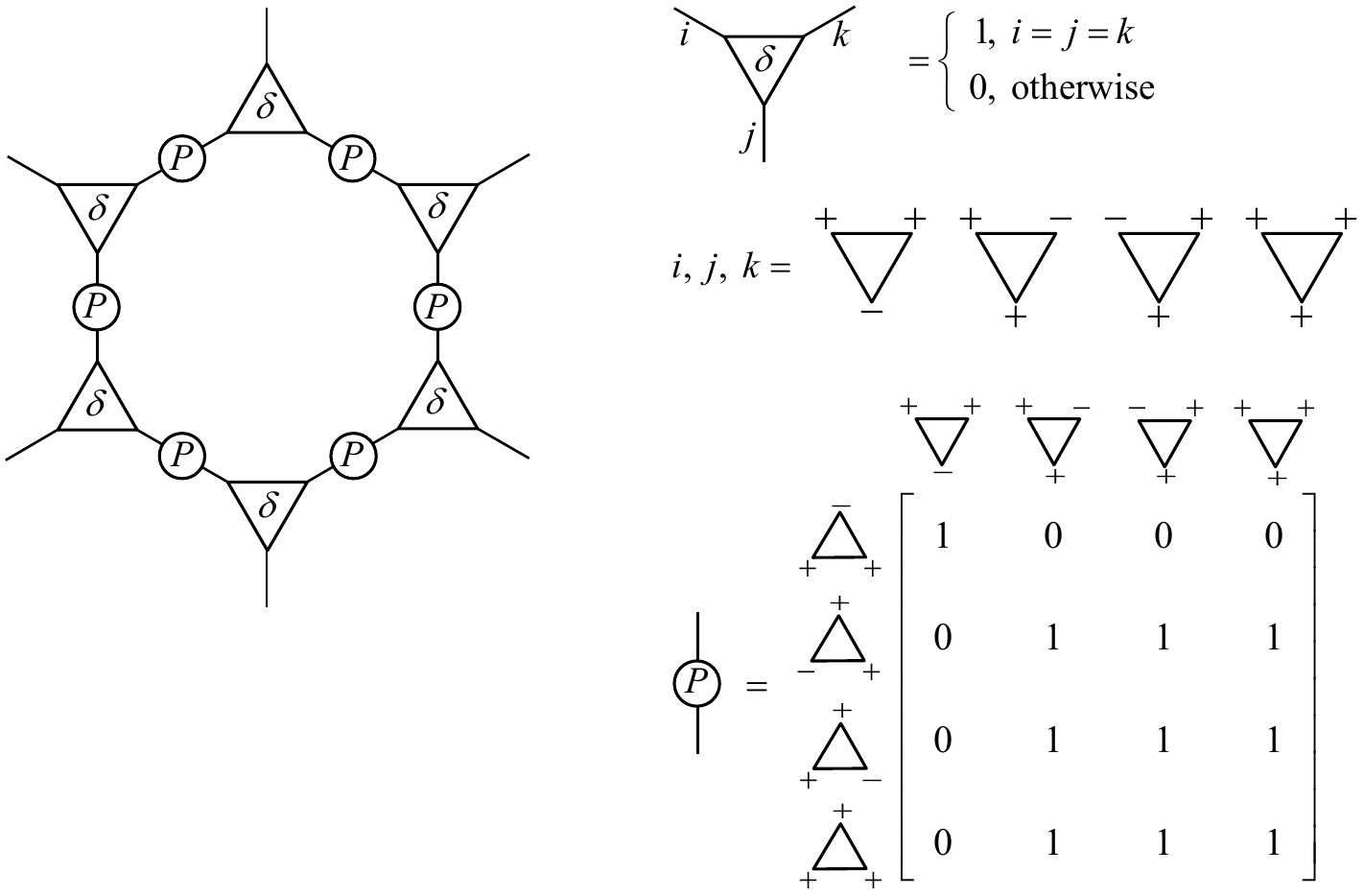} 
        \caption{The tensor network construction depending on the local rule for the Kagom\'e Ising ground states in the critical field. The definitions of the $\delta$ tensor and the $P$ matrix are explicitly presented.}  \label{fig2}
\end{figure}

As we have analysed in Sec.~\ref{Ising}, the local ground states of a Kagom\'e triangle in the critical field are $2 \times \left(+1\right) + 1 \times \left(-1\right)$ and $3 \times \left(+1\right)$. Therefore, the tensor network can be defined using the local tensors for the four ground states and the selecting matrices that enforce consistency of the spins shared by nearest-neighbour triangles. To be concrete, in the center of each triangle we set a $\delta$ tensor with rank 3 and bond dimension 4 representing the four local ground state configurations. At the position of each spin, we place a $P$ matrix with bond dimension 4, whose elements equal 1 when the two connected configurations have the same value of this shared spin and 0 otherwise. We show the tensor network as well as the definitions of the $\delta$ and $P$ tensors in Figure \ref{fig2}. Although our construction differs from that of Refs.~\cite{RN181, RN345} in the local ground states, the tensor network is well-defined and the contraction produces the residual entropy in the critical field.

\begin{figure}
 \centering
        \includegraphics[width=0.75\textwidth]{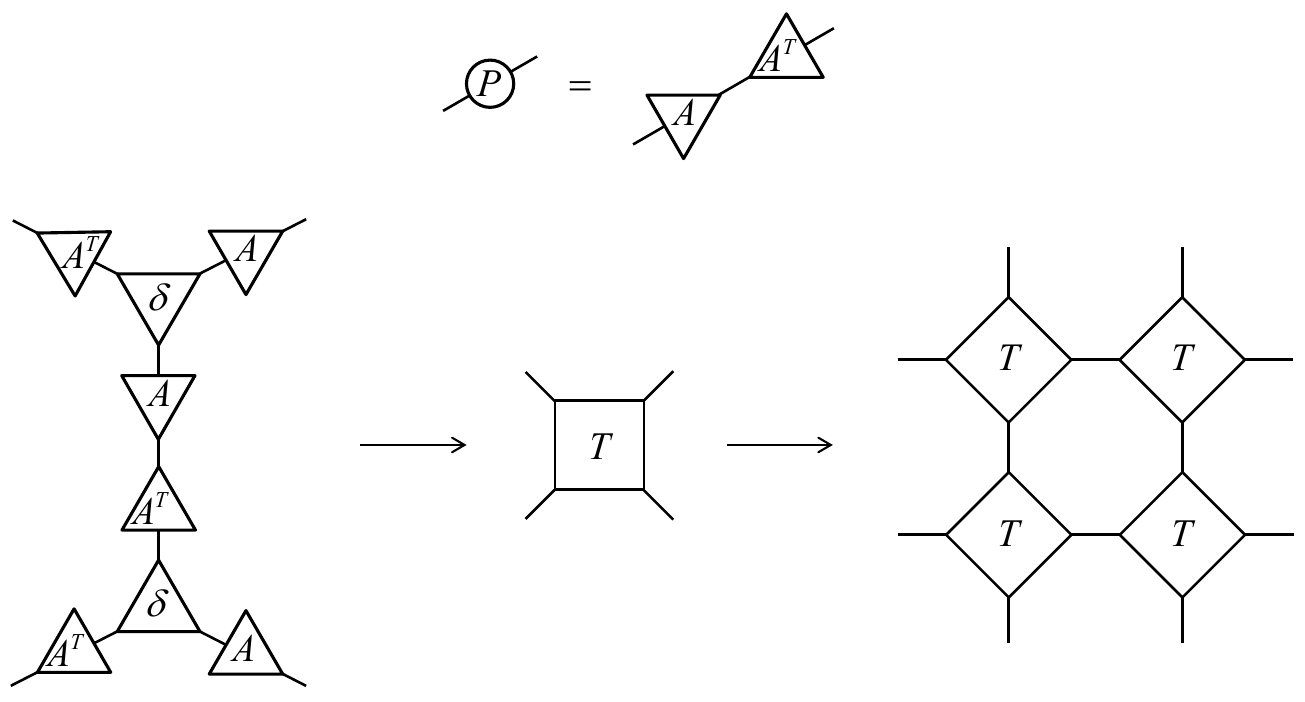} 
        \caption{The tensor network in Figure \ref{fig2} is transformed into the $T$ tensors on the square lattice. This square lattice tensor network is obtained by performing an SVD on the $P$ matrices and contracting the internal indices of each unit consisting of two nearest-neighbour Ising triangles.}  \label{fig3}
\end{figure}

As suggested by Ref.~\cite{RN181}, we perform a singular value decomposition (SVD) on the $P$ matrices to reduce the bond dimension. Explicitly, the SVD is expressed as
\begin{equation}
P = A A^T,    \label{eq2}
\end{equation}
with
\begin{equation}
A = \left[ {\begin{array}{*{20}{c}} 1&0 \\ 0&1 \\ 0&1 \\ 0&1 \end{array}} \right].  \label{eq3}
\end{equation}
By regarding two nearest-neighbour Ising triangles (two nearest-neighbour sites on the honeycomb lattice) as a unit and contracting the internal indices of each unit, we obtain the $T$ tensors on the square lattice, as shown in Figure \ref{fig3}. Each $T$ tensor is of rank 4 and bond dimension 2. One can find that, the tensor network formed by the $T$ tensors on the square lattice is equivalent to the sixteen-vertex model discussed in Sec.~\ref{summary}, and the 16 elements of each $T$ tensor coincide with the vertex weights listed in Eq.~(\ref{eq19}) for $z_1=z_2=z_3=z=1$.

\subsection{Numerical results}
We contract the infinite square-lattice tensor network by using the variational uniform matrix product state (VUMPS) algorithm \cite{RN351, RN349, RN352, RN594}. To make the numerical procedure explicit, we denote the local rank-4 tensor obtained in Sec.~\ref{constr} by $T_{lr}^{ud}$. The indices $l,r=1,\ldots,p$ are the left and right virtual indices on the square lattice, while $u,d=1,\ldots,q$ are the upper and lower indices. For the tensor constructed in Figure~\ref{fig3}, both dimensions are $p=q=2$. An infinite horizontal row of $T$ tensors is then regarded as a matrix product operator (MPO), denoted by $\mathcal{T}$. Its dominant boundary fixed point is approximated by a one-site uniform matrix product state (MPS),
\begin{equation}
|\Psi(A)\rangle=\sum_{\{s_n\}}\mathop{\rm tr}\left(\cdots A^{s_{n-1}}A^{s_n}A^{s_{n+1}}\cdots\right)|\ldots s_{n-1}s_ns_{n+1}\ldots\rangle . \label{eq4}
\end{equation}
Here $A^s$ is a $D\times D$ matrix, or equivalently $A_{asb}$ is a rank-3 tensor of shape $(D,q,D)$. The index $s=1,\ldots,q$ is connected to one vertical leg of the MPO, the indices $a,b=1,\ldots,D$ are the left and right MPS virtual indices, and $D$ is the boundary-MPS bond dimension controlling the numerical accuracy.

We represent the MPS in mixed canonical form by four tensors: the left-isometric tensor $A_L$, the right-isometric tensor $A_R$, the bond-center matrix $C$, and the one-site center tensor $A_C$. For each physical index $s$, they obey
\begin{equation}
A_C^s=A_L^sC=CA_R^s, \label{eq5}
\end{equation}
together with the isometric conditions
\begin{equation}
\sum_s (A_L^s)^\dagger A_L^s=I_D, \qquad
\sum_s A_R^s(A_R^s)^\dagger=I_D. \label{eq6}
\end{equation}
Thus $A_L$ and $A_R$ describe the semi-infinite left and right parts of the boundary MPS, respectively; $C$ contains the Schmidt coefficients across the chosen bond; and $A_C$ is the variational tensor at the center site. We normalize $C$ by its Frobenius norm. In the initial step, a random complex tensor $A$ is transformed into this form by alternating QR/RQ factorizations with matrix-free dominant-eigenvector calculations.

For fixed $A_L$ and $A_R$, the contraction of a semi-infinite part of the MPS--MPO--MPS network is represented by the left and right environment tensors $F_L$ and $F_R$. Explicitly, $F_L(l,a,\bar a)$ and $F_R(r,b,\bar b)$ have shape $(W,D,D)$; $l,r$ are MPO virtual indices, $a,b$ belong to the ket MPS, and $\bar a,\bar b$ belong to its complex-conjugate bra. Let $\mathcal{E}_L$ and $\mathcal{E}_R$ denote the linear maps obtained by adding one local tensor $T$, one MPS tensor, and one conjugate MPS tensor to the corresponding environment. The environments satisfy the largest-modulus fixed-point equations
\begin{equation}
\mathcal{E}_L(F_L)=\lambda_LF_L, \qquad
\mathcal{E}_R(F_R)=\lambda_RF_R. \label{eq7}
\end{equation}
Both equations are solved by the Arnoldi method without explicitly constructing the matrices of dimension $WD^2\times WD^2$. We take $\lambda_D=(\lambda_L+\lambda_R)/2$ as the dominant eigenvalue per square-lattice tensor and fix the remaining relative normalization of the environments by
\begin{equation}
\sum_{l,a,\bar a,b,\bar b}F_L(l,a,\bar a)C_{ab}\bar C_{\bar a\bar b}F_R(l,b,\bar b)=1. \label{eq8}
\end{equation}

With $F_L$ and $F_R$ fixed, all tensors outside the center site can be contracted. This defines the effective one-site map $\mathcal{O}_{A_C}$, acting on a trial center tensor of shape $(D,q,D)$, and the effective zero-site map $\mathcal{O}_C$, acting on a trial $D\times D$ bond matrix. The former contains $F_L$, one local MPO tensor $T$, and $F_R$, whereas the latter joins the two environments without a physical site. One VUMPS iteration is performed as follows:
\begin{enumerate}
\item Solve Eq.~(\ref{eq7}) for $F_L$, $F_R$, $\lambda_L$, and $\lambda_R$, using the environments from the preceding iteration as initial vectors.
\item Solve the two local largest-modulus eigenvalue equations
\begin{equation}
\mathcal{O}_{A_C}(A_C')=\mu_{A_C}A_C', \qquad
\mathcal{O}_C(C')=\mu_CC', \label{eq9}
\end{equation}
where $\mu_{A_C}$ and $\mu_C$ are the corresponding effective eigenvalues, and the primes denote the newly optimized center tensors.
\item Remove the arbitrary relative phase between $A_C'$ and $C'$. New isometric tensors $A_L'$ and $A_R'$ are then obtained from separate left and right polar decompositions such that $A_C'\simeq A_L'C'$ and $A_C'\simeq C'A_R'$. This construction avoids the inverse of $C'$ and remains stable when some Schmidt coefficients are small.
\item Replace $(A_L,A_R,A_C,C)$ by $(A_L',A_R',A_C',C')$ and repeat the calculation until convergence.
\end{enumerate}
The main convergence quantity reported below is the tangent-space residual
\begin{equation}
\epsilon_D=\frac{\|\mathcal{O}_{A_C}(A_C)-A_L\mathcal{O}_C(C)\|_F}
{\max\!\left[\|\mathcal{O}_{A_C}(A_C)\|_F,\|A_L\mathcal{O}_C(C)\|_F\right]}, \label{eq10}
\end{equation}
where $(A_LX)^s=A_L^sX$ and $\|\cdot\|_F$ denotes the Frobenius norm. We additionally monitor the relative eigenvalue mismatch $|\lambda_L-\lambda_R|/\max(1,|\lambda_L|,|\lambda_R|)$, the reconstruction errors of $A_C-A_LC$ and $A_C-CA_R$, and the center compatibility error of $A_LC-CA_R$. The iteration is stopped only when the maximum of these quantities is smaller than $10^{-13}$. The canonicalization and all inner Arnoldi eigenvalue problems are solved with the stricter tolerance $10^{-14}$.

To examine the dependence on the boundary-MPS bond dimension, we perform independent calculations for $D=20$, 25, 30, 35, and 40. The results are listed in Table~\ref{tab1}. The small imaginary part of $\lambda_D$, whose magnitude is below $5\times10^{-14}$ for all calculations, originates from numerical round-off. We therefore use $\mathrm{Re}\,\lambda_D$ in evaluating the entropy.

\begin{table}
\caption{Convergence of the VUMPS contraction with the boundary-MPS bond dimension $D$. The residual entropy per honeycomb-lattice site is $S_{\rm hon}=\frac{1}{2}\ln(\mathrm{Re}\,\lambda_D)$, and the corresponding value per Kagom\'e-lattice site is $S_{\rm Kag}=\frac{1}{3}\ln(\mathrm{Re}\,\lambda_D)$.}
\centering
\vspace{1mm}
\begin{tabular}{c c c c c}
\hline
$D$ & $\mathrm{Re}\,\lambda_D$ & $S_{\rm hon}$ & $S_{\rm Kag}$ & $\epsilon_D$ \\
\hline
20 & 3.200799853397746 & 0.581700366379567 & 0.387800244253044 & $1.74\times10^{-14}$ \\
25 & 3.200799853397769 & 0.581700366379570 & 0.387800244253047 & $1.10\times10^{-14}$ \\
30 & 3.200799853397762 & 0.581700366379569 & 0.387800244253046 & $4.49\times10^{-14}$ \\
35 & 3.200799853397763 & 0.581700366379569 & 0.387800244253046 & $5.03\times10^{-14}$ \\
40 & 3.200799853397763 & 0.581700366379569 & 0.387800244253046 & $6.74\times10^{-14}$ \\
\hline
\end{tabular}  \label{tab1}
\end{table}

Each $T$ tensor contains two sites of the original honeycomb lattice. Hence, in the thermodynamic limit, the solution of the monomer-dimer model is related to the dominant eigenvalue by
\begin{equation}
\mathop {\lim}\limits_{N \to \infty}\frac{1}{N}\ln Z
=\frac{1}{2}\ln\left(\mathrm{Re}\,\lambda_D\right)
=0.581700366379569. \label{eq11}
\end{equation}
The number of Kagom\'e-lattice sites is $\frac{3}{2}N$. The residual entropy per Kagom\'e-lattice site at the critical field is therefore
\begin{equation}
S_{\rm Kag}=\frac{2}{3}\mathop {\lim}\limits_{N \to \infty}\frac{1}{N}\ln Z
=\frac{1}{3}\ln\left(\mathrm{Re}\,\lambda_D\right)
=0.387800244253046. \label{eq12}
\end{equation}
The values at $D=35$ and 40 agree in all the digits shown. Moreover, over the complete range $D=20$--40, the variation of $\mathrm{Re}\,\lambda_D$ is only $2.4\times10^{-14}$ and that of $S_{\rm Kag}$ is $2.5\times10^{-15}$. These results demonstrate that the contraction has already saturated with respect to the boundary-MPS bond dimension at the accuracy considered here.

\subsection{Comparison with previous estimates}
Table \ref{tab2} lists the estimates for the solution of the monomer-dimer model on the honeycomb lattice from previous studies. Some of the estimates are obtained from the studies on the Kagom\'e Ising model and spin ice model, and we have transformed them into the values of the equivalent monomer-dimer problem. We briefly review these results. The Bethe approximation for the monomer-dimer model was studied by Nagle \cite{RN147}. The solution when the dimer and monomer activities are 1 is given in Eq.~(39) of Ref.~\cite{RN147} 
\begin{equation}
\mathop {\lim}\limits_{N \to \infty}\frac{1}{N}\ln Z = \ln z + \frac{n}{2}\ln \frac{(n-1)\left[1 + \sqrt{1 + \frac{4(n-1)\tilde z}{z^2}} \right]}{n - 2 + n\sqrt{1 + \frac{4(n-1)\tilde z}{z^2}}} + \ln \frac{n - 2 + n\sqrt{1 + \frac{4(n-1)\tilde z}{z^2}}}{2(n-1)}~,   \label{eq13} 
\end{equation}
where $n$ is the coordination number of each site on the lattice and $\tilde z$ and $z$ are the dimer and monomer activities, respectively. For the honeycomb lattice, we take $n=3$ and $\tilde z = z = 1$ and obtain the estimate of Bethe approximation $\frac{1}{2} \ln \left( \frac{16}{5} \right)$. Isakov \textit{et al.} \cite{RN104} rederived the Bethe approximation in a different way, giving the same result. The Wang-Landau estimate in Ref.~\cite{RN120} was obtained from the sampling performed on the Kagom\'e Ising model in the critical field. The value $0.582009 \pm 0.000023$ is larger than the Bethe approximation result. Estimates based on Monte Carlo simulation \cite{RN567} and tensor network calculation \cite{RN354} have also been reported, which are very close and slightly larger than the Bethe approximation result.

\begin{table}
\caption{The estimates for the solution of the monomer-dimer model on the honeycomb lattice, and comparison with our result.}
\centering
\vspace{1mm}
\begin{tabular}{l l l}
\hline
Group & Method & Estimate \\
\hline
Nagle \cite{RN147} & Bethe approximation & $\frac{1}{2} \ln \left( \frac{16}{5} \right) = 0.5815754$ \\
Isakov \textit{et al.}~\cite{RN104} & Bethe approximation & same as above \\
Andriushchenko \textit{et al.}~\cite{RN120} & Wang-Landau sampling & $0.582009 \pm 0.000023$ \\
Semjan \textit{et al.}~\cite{RN567} & Monte Carlo simulation & 0.5817 \\
Colbois \textit{et al.}~\cite{RN354} & tensor network (VUMPS) & $0.581700366380 \pm 10^{-12}$ \\
Our result & tensor network (VUMPS) & $0.581700366379569 \pm 10^{-15}$ \\
\hline
\end{tabular}
\label{tab2}
\end{table}
%
%

It is seen clearly that our result is in excellent agreement with that of Ref.~\cite{RN354}, which is obtained using a tensor network construction slightly different from ours (the authors of Ref.~\cite{RN354} had also examined their result by using a tensor network similar to ours, see the footnote [77] therein). The additional stable digits primarily result from the larger boundary bond dimension used here---we use $D=40$ as shown in Table \ref{tab1} and the largest $D$ for the calculation in Ref.~\cite{RN354} is 10. This comparison confirms the high precision of our calculation. 

To further examine our method, we perform numerical calculations for monomer activities $z=0.2,~0.4,~0.6,~0.8$ and fixed dimer activity $\tilde z=1$. We slightly modify the tensor network construction in Figures \ref{fig2} and \ref{fig3} by replacing the $\delta$ tensor with a weighted $\delta$ tensor
\begin{equation}
\tilde \delta_{ijk} = \left\{ \begin{array}{*{20}{l}}
1,~i=j=k=2 \times \left(+1\right) + 1 \times \left(-1\right) \\
z,~i=j=k=3 \times \left(+1\right) \\
0,~\rm{otherwise}
\end{array} \right. .   \label{eqdelta} 
\end{equation}
Following the precedure shown in Figure \ref{fig3} we obtain the tensor network formed by the $\tilde T$ tensors on the square lattice, which is equivalent to the sixteen-vertex model discussed in Sec.~\ref{summary}. The 16 elements of each $\tilde T$ tensor coincide with the vertex weights listed in Eq.~(\ref{eq19}) for $z_1=z_2=z_3=1$. The VUMPS algorithm is again employed to contract the tensor network with boundary bond dimension $D=40$. As in the case that $z=1$, the numerical values are obtained using a convergence tolerance of $10^{-15}$. Table \ref{tab3} lists our values and the Bethe approximation results, where the Bethe approximation results are obtained from Eq.~(\ref{eq13}). It is seen that, for each $z$, our estimate is slightly larger than Bethe approximation result. The comparison illustrates that our estimates are in good agreement with the theoretical approximations, and our method is effective for the monomer-dimer problem. We note that the solutions with monomer activities below 0.1 have been studied in Ref.~\cite{RN347}, using the higher-order tensor renormalization group (HOTRG) method \cite{RN694} and density matrix renormalization group (DMRG) method \cite{RN341, RN768}.

\begin{table}
\caption{Our estimates for the solutions of the monomer-dimer model on the honeycomb lattice with monomer activities $z=0.2,~0.4,~0.6,~0.8,~1$ and fixed dimer activity $\tilde z=1$, and comparison with the Bethe approximation results. Our numerical values are obtained using a convergence tolerance of $10^{-15}$. The reported values are the final converged results of the algorithm.}
\centering
\vspace{1mm}
\begin{tabular}{l l l}
\hline
$z$ & Bethe approximation & Our result \\ 
\hline
\vspace{1mm} 0.2 & $\frac{1}{2} \ln \left[ \frac{2(151+51\sqrt{201})}{25(1+3\sqrt{201})} \right] = 0.2369523$ & 0.241536357806263 \\
\vspace{1mm} 0.4 & $\frac{1}{2} \ln \left[ \frac{4(77+27\sqrt{51})}{25(1+3\sqrt{51})} \right] = 0.3275119$ & 0.329194419749283 \\ 
\vspace{1mm} 0.6 & $\frac{1}{2} \ln \left[ \frac{2(477+59\sqrt{209})}{75(1+\sqrt{209})} \right] = 0.4152535$ & 0.415930411091480 \\
\vspace{1mm} 0.8 & $\frac{1}{2} \ln \left[ \frac{4(166+99\sqrt{6})}{25(2+9\sqrt{6})} \right] = 0.4999826$ & 0.500268633085667 \\
1 & $\frac{1}{2} \ln \left( \frac{16}{5} \right) = 0.5815754$ & 0.581700366379569 \\
\hline
\end{tabular}
\label{tab3}
\end{table}

\section{Edge coloring problem of the honeycomb lattice}   \label{coloring}
We consider in this section the problem of enumeration of colorings for the edges of the honeycomb lattice \cite{RN358, RN360, RN533}. It is natural to relate the pure dimer model and the monomer-dimer model to the edge coloring problem, as the edge covered/uncovered by a dimer can be seen as being colored with color $A$/$B$. One can also use three or more colors, which can lead to the configurations of the multi-state vertex model. Therefore, it is beneficial to formulate the monomer-dimer model as well as other relevant problems in the language of edge coloring. Here we revisit some cases of the edge coloring problem of the honeycomb lattice using two or three colors, which are termed two- or three-coloring, respectively. 

Consider the three edges incident to a site. We use the notation $XXX$ to represent the local coloring rules. For example, the $AAB$ case of two-coloring is that the three edges incident to each site are colored in the way---two edges are colored with $A$ and one is colored with $B$. Denote the number of colorings by $W$. We list the results for some cases of two-coloring as follows: \\
\noindent (\romannumeral1). $AAB$. Since each edge colored with $B$ can be seen as being covered by a dimer, we can verify that this case is equivalent to the pure dimer model on the honeycomb lattice. The exact solution is \cite{RN140}
\begin{equation}
\mathop {\lim}\limits_{N \to \infty} \frac{1}{N}\ln W_{AAB} = \frac{1}{16 \pi ^2} \int_0^{2\pi} d\theta \int_0^{2\pi} d\phi \ln \left[ 3 + 2\left( \cos \theta + \cos \phi + \cos \left( \theta + \phi  \right) \right) \right] = 0.161533.   \label{eq14}
\end{equation}
It has been shown \cite{RN241, RN157, RN145, RN225} that the solution of the pure dimer model on the honeycomb lattice is one half of the residual entropy of antiferromagnetic triangular Ising model \cite{RN81}. We also note that $W_{AAB}$ can be expressed in another way. If we draw $m$ nonoverlapping polygons to cover the lattice, and color the edges on these polygons with $A$ and the remaining edges with $B$, we can obtain a configuration of $AAB$ coloring. Following the appendix of Ref.~\cite{RN358} we denote the number of ways of covering the lattice with $m$ nonoverlapping polygons by $g(m)$. Then we have $W_{AAB} = \sum\limits_m {g(m)}$. \\
\noindent (\romannumeral2). $AAB\&ABB$. The local coloring configurations can be mapped into the spin states $2 \times \left(+1\right) + 1 \times \left(-1\right)$ and $2 \times \left(-1\right) + 1 \times \left(+1\right)$ of a triangle, i.e., the local ground states of the antiferromagnetic Kagom\'e Ising model in the zero field. The exact solution is the residual entropy of the antiferromagnetic Kagom\'e Ising model \cite{RN82}
\begin{equation}
\mathop {\lim}\limits_{N \to \infty} \frac{1}{N}\ln W_{AAB\& ABB} = \frac{1}{16 \pi ^2} \int_0^{2\pi} d\theta \int_0^{2\pi} d\phi \ln \left[ 21 - 4\left( \cos \theta + \cos \phi + \cos \left( \theta + \phi \right) \right) \right] = 0.752745.  \label{eq15}
\end{equation}
\noindent (\romannumeral3). $AAA\&AAB$. We map each edge colored with $B$ into a dimer as in case (\romannumeral1). It is obvious that this case is equivalent to the monomer-dimer model on the honeycomb lattice. Our numerical result for this case is given in Eq.~(\ref{eq11})
\begin{equation}
\mathop {\lim}\limits_{N \to \infty} \frac{1}{N}\ln W_{AAA\& AAB} = 0.581700366379569.  \label{eq16}
\end{equation}
\noindent (\romannumeral4). $AAA\&ABB$. This problem can be seen as the case of $a=c=1, b=d=0$ studied in Ref.~\cite{RN117}. Hence, the exact solution is obtained by substituting $a=c=1$ into Eq.~(17) of Ref.~\cite{RN117}
\begin{equation}
\mathop {\lim}\limits_{N \to \infty} \frac{1}{N}\ln W_{AAA\& ABB} = \frac{1}{2} \ln 2 = 0.346574.  \label{eq17}
\end{equation}
We also find that $W_{AAA\&ABB} = \sum\limits_m h(m)$, where $h(m)$ is the number of ways of drawing $m$ nonoverlapping polygons on the lattice. This identity follows from the following one-to-one correspondence: if we draw nonoverlapping polygons on the lattice and color the edges on these polygons with $B$ and the remaining edges with $A$, we can establish a one-to-one correspondence between the ways of drawing and the $AAA\&ABB$ colorings.

For the three-coloring problem, we list one case: \\
\noindent (\romannumeral1). $ABC$. This case can be mapped into the ground state of the three-state Potts model on the Kagom\'e lattice \cite{RN212}. The exact solution has been derived by the transfer matrix method \cite{RN358}
\begin{equation}
\mathop {\lim}\limits_{N \to \infty} \frac{1}{N}\ln W_{ABC} = \frac{1}{2} \ln \left[ \prod\limits_{p = 1}^\infty \frac{{\left(3p - 1\right)}^2}{\left(3p - 2\right)3p} \right] = 0.189560.    \label{eq18}
\end{equation}
It has also been proved that $W_{ABC} = \sum\limits_m {2^m g(m)}$ in the appendix of Ref.~\cite{RN358}, where $g(m)$ again denotes the number of ways of covering the lattice with $m$ nonoverlapping polygons.

\section{Discussion and Summary}   \label{summary}
In this paper we present a tensor network study of the monomer-dimer model on the honeycomb lattice. The monomer-dimer configurations can be exactly mapped into the ground states of the antiferromagnetic Kagom\'e Ising model in the critical field. The tensor network can therefore be constructed based on the local rule for the ground states. A high-precision result in the thermodynamic limit is obtained by contracting the tensor network using the VUMPS algorithm. We compare our result with the previous estimates, including theoretical approximations and numerical calculations, and show that they are in good agreement. We also connect the monomer-dimer model to the problem of enumeration of edge colorings of the honeycomb lattice. Some cases of the edge coloring problem of the honeycomb lattice using two or three colors are revisited, with the monomer-dimer model appearing as a special case. Hence, our work also provides an estimate for one case of the edge coloring problem.

\begin{figure}
 \centering
        \includegraphics[width=0.83\textwidth]{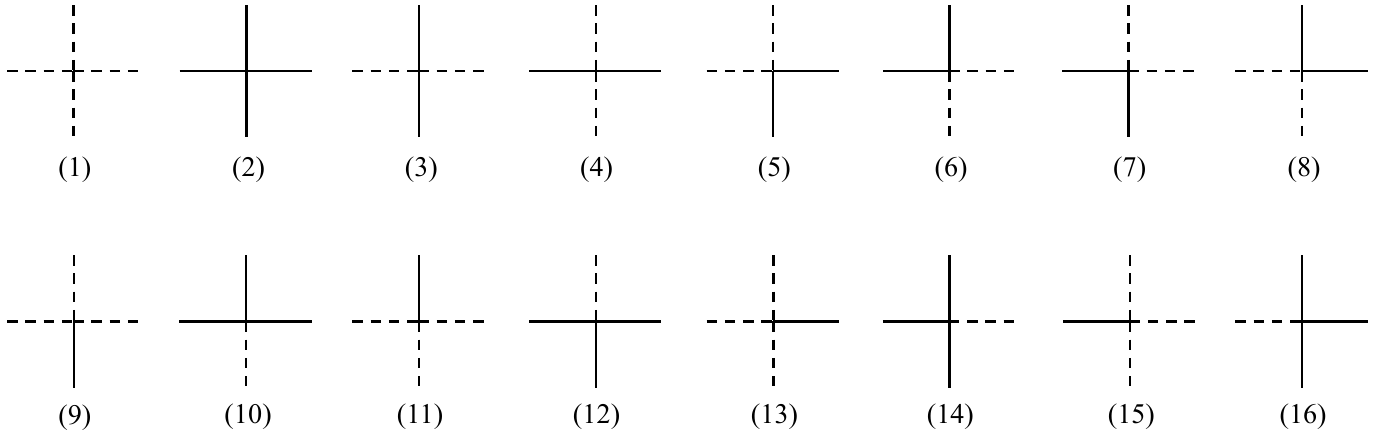}
        \caption{The vertex configurations of the sixteen-vertex model. Each solid line represents a bonded edge, while each dashed line represents an unbonded edge. The even eight-vertex model consists of vertices (1)--(8), while the odd subcase consists of vertices (9)--(16).}  \label{figsum1}
\end{figure}

\begin{figure}
 \centering
        \includegraphics[width=0.27\textwidth]{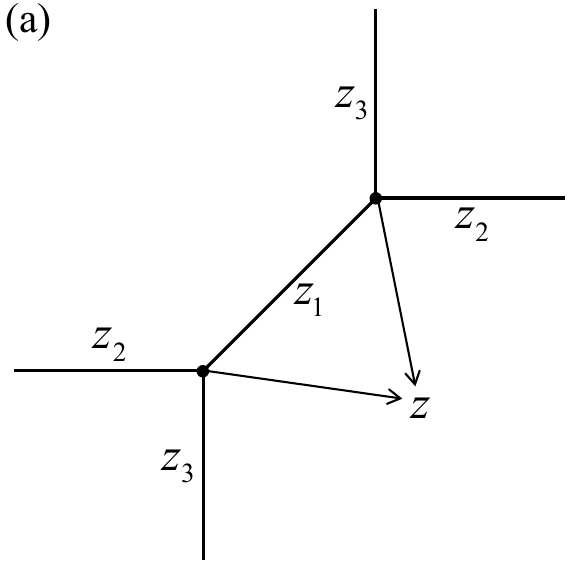}  \quad \qquad
        \includegraphics[width=0.558\textwidth]{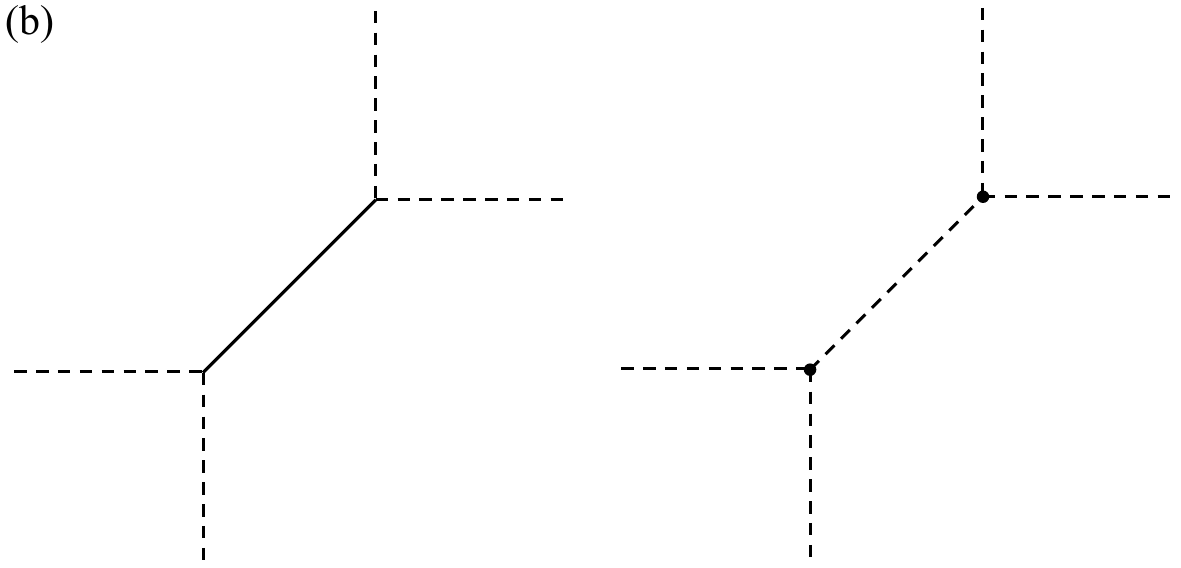}
        \caption{(a) The vertex unit of the monomer-dimer model on the honeycomb lattice with the associated dimer and monomer activities. (b) Two monomer-dimer arrangements of the unit corresponding to vertex (1) of the sixteen-vertex model in Figure \ref{figsum1}.}  \label{figsum2}
\end{figure}

As mentioned in Sec.~\ref{intro}, the monomer-dimer model is very attractive since it is a well-known unsolved problem in the field of statistical lattice model. It is of interest to us to further discuss the non-integrability of the monomer-dimer model in the language of vertex models. We consider mapping the monomer-dimer model on the honeycomb lattice into the sixteen-vertex model on the square lattice. The 16 vertex configurations are shown in Figure \ref{figsum1}. We adopt the notation of vertex configurations in Figure 1 of Ref.~\cite{RN124}, where each edge is either bonded or unbonded. The vertex unit on the honeycomb lattice, together with the associated dimer and monomer activities ($z_1$, $z_2$, $z_3$ and $z$), is shown in Figure \ref{figsum2}(a). It is straightforward to verify that the vertex units form a square lattice, thus we can use a vertex unit to represent a vertex site in Figure \ref{figsum1}. The mapping into the equivalent sixteen-vertex model can be established by defining the state of each edge connecting neighbouring vertex units (there are four such edges in each vertex unit) as follows: an edge covered by a dimer is regarded as bonded, whereas an uncovered edge is regarded as unbonded. In this way the monomer-dimer arrangements of the vertex unit in Figure \ref{figsum2}(a) can be transformed into the vertex configurations in Figure \ref{figsum1}. For example, two arrangements shown in Figure \ref{figsum2}(b) correspond to vertex (1). Then we can obtain all vertex weights of the equivalent sixteen-vertex model, as listed below
\begin{eqnarray}
&{\omega _1} = z_1 + z^2,~{\omega _2} = 0,~{\omega _3} = z_3,~{\omega _4} = z_2,~{\omega _5} = {\omega _6} = \sqrt{z_2 z_3},~{\omega _7} = {\omega _8} = 0,  \nonumber  \\
&{\omega _9} = {\omega _{11}} = \sqrt{z_3}z,~{\omega _{13}} = {\omega _{15}} = \sqrt{z_2}z,~{\omega _{10}} = {\omega _{12}} = {\omega _{14}} = {\omega _{16}} = 0.   \label{eq19}
\end{eqnarray}
This is an unsolved sixteen-vertex model to date \cite{RN124}. Readers interested in the integrability of the sixteen-vertex model can see Ref.~\cite{RN124} for a review. When $z=0$ the model reduces to the pure dimer model, and the equivalent sixteen-vertex model reduces to an exactly solvable subcase known as the even free-fermion model \cite{RN65, RN59}. The pure dimer model can thereby be solved in this way \cite{RN140}. In the Appendix, we give a free-fermion formulation for the pure dimer models on some typical two-dimensional lattices, specifically the square, honeycomb, Kagom\'e and triangular lattices.  

\ack{We thank Dr.~Jeanne Colbois, Dr.~Wei-Jie Huang and Mr.~Tong-Yu Lin for discussions, and Prof.~Petr Andriushchenko for giving us the Wang-Landau result in his paper.}

\funding{This work was supported by National Natural Science Foundation of China (Grants No.~12605080 and No.~12474489), Guangdong Provincial Quantum Science Strategic Initiative (Grants No.~GDZX2203001 and No.~GDZX2403001), Shenzhen Fundamental Research Program (Grant No.~JCYJ20240813153139050), and Research Funding for Outbound Postdoctoral Fellows in Shenzhen (Grant No.~SZRCXM2401006).}

\roles{Conceptualization, D.-Z.L.; methodology, D.-Z.L.; validation, D.-Z.L., J.L.~and X.W.; formal analysis, D.-Z.L.~and J.L.; investigation, D.-Z.L.~and J.L.; resources, J.L.; software, J.L.; data curation, J.L.; visualization, D.-Z.L.; writing--original draft, D.-Z.L.~and J.L.; writing--review and editing, X.W.; supervision, X.W.; project administration, X.W.; funding acquisition, X.W.~and D.-Z.L.}

\data{All data that support the findings of this study are included within the article.}

\appendix
\section*{Appendix. Free-fermion formulation of the two-dimensional pure dimer models}   \label{appendix}
In this appendix, we present the free-fermion formulation for the pure dimer models on the square, honeycomb, Kagom\'e and triangular lattices. As shown in Figure \ref{figsum1}, the sixteen-vertex model includes two subcases: the even eight-vertex model [vertices (1)--(8)] and the odd eight-vertex model [vertices (9)--(16)]. The free-fermion eight-vertex model is defined by the well-known free-fermion condition, which is either 
\begin{equation}
\omega _1\omega _2 + \omega _3\omega _4 = \omega _5\omega _6 + \omega _7\omega _8   \label{eqa1}
\end{equation}
for the even subcase \cite{RN65, RN59}, or
\begin{equation}
\omega _9\omega _{10} + \omega _{11}\omega _{12} = \omega _{13}\omega _{14} + \omega _{15}\omega _{16}   \label{eqa2}
\end{equation}
for the odd subcase \cite{RN129}. Both even and odd free-fermion models are exactly solvable. We quote the corresponding solutions from Refs.~\cite{RN59} and \cite{RN129}:
\begin{equation}
\mathop {\lim}\limits_{N_{\rm{site}} \to \infty } \frac{1}{N_{\rm{site}}} \ln Z_{\rm{even}} = \frac{1}{8 \pi ^2}\int_0^{2\pi} d\theta \int_0^{2\pi} d\phi \ln \left[ a + b\cos \theta + c\cos \phi + d\cos \left( \theta - \phi \right) + e\cos \left( \theta + \phi \right) \right]   \label{eqa3}
\end{equation}
with
\begin{eqnarray}
&a = \omega _1^2 + \omega _2^2 + \omega _3^2 + \omega _4^2,  \nonumber \\
&b = 2\left( \omega_1 \omega_3 - \omega_2 \omega_4 \right),  \nonumber \\
&c = 2\left( \omega_1 \omega_4 - \omega_2 \omega_3 \right), \nonumber \\
&d = 2\left( \omega_3 \omega_4 - \omega_7 \omega_8 \right), \nonumber \\
&e = 2\left( \omega_3 \omega_4 - \omega_5 \omega_6 \right),  \label{eqa4}
\end{eqnarray}
for the even free-fermion model; and 
\begin{equation}
\mathop {\lim}\limits_{N_{\rm{site}} \to \infty } \frac{1}{N_{\rm{site}}} \ln Z_{\rm{odd}} = \frac{1}{16 \pi ^2}\int_0^{2\pi} d\theta \int_0^{2\pi} d\phi \ln \left[ \tilde a + \tilde b\cos \theta + \tilde c\cos \phi + \tilde d\cos \left( \theta - \phi \right) + \tilde e\cos \left( \theta + \phi \right) \right]   \label{eqa5}
\end{equation}
with
\begin{eqnarray}
&\tilde a = 2 \left[ \left( \omega_9 \omega_{10} + \omega_{11} \omega_{12} \right)^2 + \left( \omega_9 \omega_{11} \right)^2 + \left( \omega_{10} \omega_{12} \right)^2 + \left( \omega_{13} \omega_{15} \right)^2 + \left( \omega_{14} \omega_{16} \right)^2 \right],  \nonumber \\
&\tilde b = 2\left[ -\left( \omega_9 \omega_{11} \right)^2 -\left( \omega_{10} \omega_{12} \right)^2 + 2 \omega_{13} \omega_{14} \omega_{15} \omega_{16} \right],  \nonumber \\
&\tilde c = 2\left[ \left( \omega_{13} \omega_{15} \right)^2 + \left( \omega_{14} \omega_{16} \right)^2 - 2 \omega_{9} \omega_{10} \omega_{11} \omega_{12} \right], \nonumber \\
&\tilde d = -2 \left( \omega_9 \omega_{10} - \omega_{13} \omega_{14} \right)^2, \nonumber \\
&\tilde e = -2 \left( \omega_{11} \omega_{12} - \omega_{13} \omega_{14} \right)^2,  \label{eqa6}
\end{eqnarray}
for the odd free-fermion model. Here $N_{\rm{site}}$ represents the number of vertex sites.

\begin{figure}
 \centering
        \includegraphics[width=0.2\textwidth]{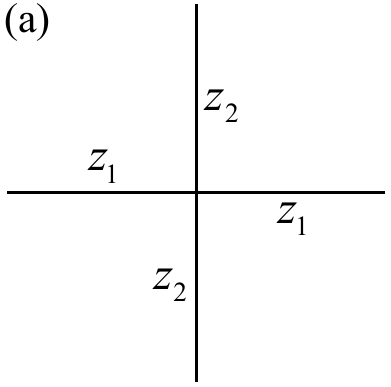}  \qquad  \qquad
        \includegraphics[width=0.3\textwidth]{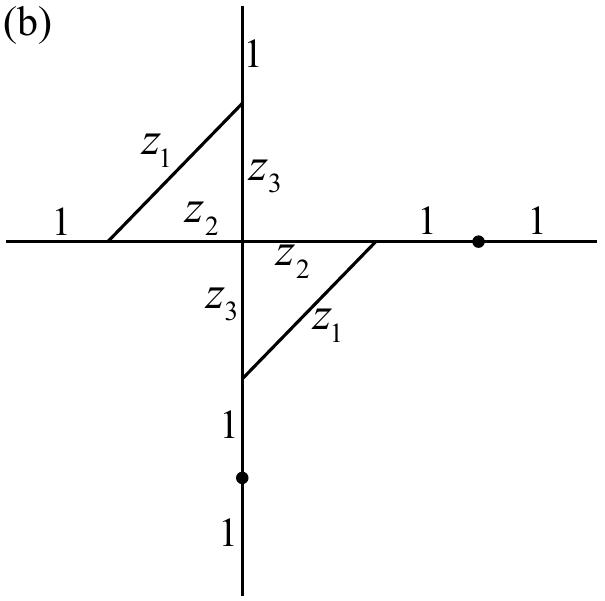}
        \caption{(a) The vertex unit of the pure dimer model on the square lattice with the associated dimer weights. (b) The vertex unit of the pure dimer model on the extended Kagom\'e lattice with the associated dimer weights.}  \label{figapp1}
\end{figure}

The pure dimer model on the square lattice is an odd free-fermion model \cite{RN129}. We show the vertex unit in Figure \ref{figapp1}(a). Using a mapping similar to that employed for the monomer-dimer model in Sec.~\ref{summary}, we can obtain the vertex weights of the equivalent sixteen-vertex model
\begin{equation}
\omega _9 = \omega _{11} = \sqrt{z_2},~\omega _{13} = \omega _{15} = \sqrt{z_1},~\rm{all~others~weights} = 0.   \label{eqa7}
\end{equation}
Obviously the odd free-fermion condition [Eq.~(\ref{eqa2})] is satisfied. Substituting Eq.~(\ref{eqa7}) into Eqs.~(\ref{eqa5}) and (\ref{eqa6}) yields the solution
\begin{equation}
\mathop {\lim}\limits_{N_{\rm{squ}} \to \infty} \frac{1}{N_{\rm{squ}}} \ln Z_{\rm{squ}} = \frac{1}{16 \pi ^2} \int_0^{2\pi} d\theta \int_0^{2\pi} d\phi \ln \left[ 2\left( z_1^2 + z_2^2 - z_2^2\cos \theta + z_1^2\cos \phi \right) \right].   \label{eqa8}
\end{equation}

The pure dimer model on the honeycomb lattice can be conveniently transformed into the even free-fermion model \cite{RN140}. The vertex unit is shown in Figure \ref{figsum2}(a) and we just need to set the monomer activity $z=0$. The vertex weights are then determined from Eq.~(\ref{eq19})
\begin{equation}
\omega _1 = z_1,~\omega _3 = z_3,~\omega _4 = z_2,~\omega _5 = \omega _6 = \sqrt{z_2 z_3},~\rm{all~others~weights} = 0.   \label{eqa9}
\end{equation}
It is straightforward to examine that the even free-fermion condition [Eq.~(\ref{eqa1})] holds in this case. Using the relation between the number of lattice points and that of the vertex sites $N_{\rm{hon}}=2N_{\rm{site}}$, we obtain the solution from Eqs.~(\ref{eqa3}) and (\ref{eqa4})
\begin{equation}
\mathop {\lim}\limits_{N_{\rm{hon}} \to \infty} \frac{1}{N_{\rm{hon}}} \ln Z_{\rm{hon}} = \frac{1}{16 \pi ^2} \int_0^{2\pi} d\theta \int_0^{2\pi} d\phi \ln \left[ z_1^2 + z_2^2 + z_3^2 + 2 z_1 z_3\cos \theta + 2 z_1 z_2\cos \phi + 2 z_2 z_3 \cos \left( \theta - \phi \right) \right].   \label{eqa10}
\end{equation}
When $z_1=z_2=z_3=1$, Eq.~(\ref{eqa10}) reduces to Eq.~(\ref{eq14}). 

The pure dimer model on the Kagom\'e lattice can be mapped into an odd free-fermion model. The mapping method, where an extended Kagom\'e lattice is constructed and the resulting dimer configurations are in one-to-one correspondence with those on the original lattice, was introduced in Ref.~\cite{RN144}. The vertex unit on the extended Kagom\'e lattice is defined in Figure 4 therein, which is also shown in Figure \ref{figapp1}(b). The vertex weights are listed 
\begin{equation}
\omega _9 = \omega _{12} = z_1 z_3,~\omega _{10} = \omega _{11} = z_2,~\omega _{13} = \omega _{16} = z_1 z_2,~\omega _{14} = \omega _{15} = z_3,~\rm{all~others~weights} = 0.   \label{eqa11}
\end{equation}
The solution is derived by substituting Eq.~(\ref{eqa11}) into Eqs.~(\ref{eqa5}) and (\ref{eqa6}) 
\begin{equation}
\mathop {\lim}\limits_{N_{\rm{Kag}} \to \infty} \frac{1}{N_{\rm{Kag}}} \ln Z_{\rm{Kag}} = \frac{1}{6} \ln \left( 4 z_1 z_2 z_3 \right) ,   \label{eqa12}
\end{equation}
which has a surprisingly simple expression.

\begin{figure}
 \centering
        \includegraphics[width=0.35\textwidth]{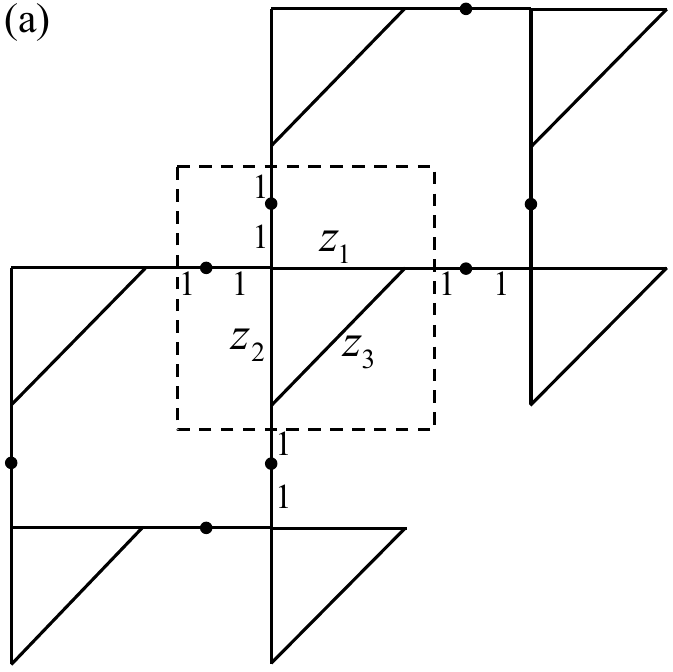}  \qquad  \qquad
        \includegraphics[width=0.18\textwidth]{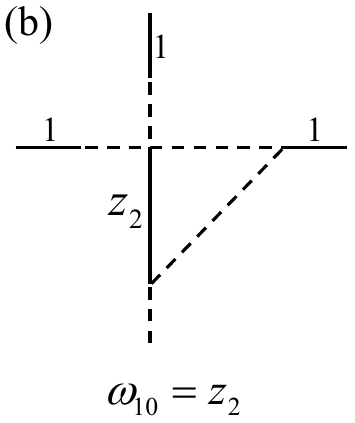}
        \caption{(a) The extended triangular lattice with the associated dimer weights. The region bounded by dashed lines forms a vertex unit. (b) An example of determining the weight of vertex (10) of the sixteen-vertex model in Figure \ref{figsum1}.}  \label{figapp2}
\end{figure}

We employ a technique similar to that introduced in Ref.~\cite{RN144} for the Kagom\'e lattice, to map the pure dimer model on the triangular lattice into the odd free-fermion model. We construct an extended triangular lattice by inserting a decorating site attached to two inserted edges of activity 1, as shown in Figure \ref{figapp2}(a). It can be verified that the dimer configurations on the extended triangular lattice are exactly equivalent to those on the original lattice. The vertex unit is the region bounded by dashed lines in Figure \ref{figapp2}(a). The vertex weights are then obtained in the same manner as that for the Kagom\'e lattice. We give an example of determining $\omega_{10}$ in Figure \ref{figapp2}(b). All vertex weights are listed
\begin{equation}
\omega _{10} = z_2,~\omega _{11} = z_3,~\omega _{12} = 1,~\omega _{14} = z_1,~\omega _{15} = z_3,~\omega _{16} = 1,~\rm{all~others~weights} = 0.   \label{eqa13}
\end{equation}
Obviously this is an odd free-fermion model. Again, the solution follows from Eqs.~(\ref{eqa5}) and (\ref{eqa6}) 
\begin{equation}
\mathop {\lim}\limits_{N_{\rm{tri}} \to \infty} \frac{1}{N_{\rm{tri}}} \ln Z_{\rm{tri}} = \frac{1}{16 \pi ^2} \int_0^{2\pi} d\theta \int_0^{2\pi} d\phi \ln \left[ 2\left( z_1^2 + z_2^2 + z_3^2 - z_2^2\cos \theta + z_1^2\cos \phi - z_3^2\cos \left( \theta + \phi \right) \right) \right].   \label{eqa14}
\end{equation}
This result is consistent with the known expressions \cite{RN143, RN145}.

\bibliographystyle{iopart-num}
\bibliography{manuscript.bib}

@article{RN406,
   author = {Allegra, Nicolas},
   title = {{Exact solution of the 2d dimer model: Corner free energy, correlation functions and combinatorics}},
   journal = {Nucl. Phys. B},
   volume = {894},
   pages = {685-732},
   ISSN = {0550-3213},
   DOI = {10.1016/j.nuclphysb.2015.03.022},
   url = {https://www.sciencedirect.com/science/article/pii/S0550321315001042},
   year = {2015},
   type = {Journal Article}
}

@article{RN580,
   author = {Allegra, Nicolas and Fortin, Jean-Yves},
   title = {{Grassmannian representation of the two-dimensional monomer-dimer model}},
   journal = {Phys. Rev. E},
   volume = {89},
   number = {6},
   pages = {062107},
   DOI = {10.1103/PhysRevE.89.062107},
   url = {https://link.aps.org/doi/10.1103/PhysRevE.89.062107},
   year = {2014},
   type = {Journal Article}
}

@article{RN120,
   author = {Andriushchenko, Petr and Soldatov, Konstantin and Peretyatko, Alexey and Shevchenko, Yuriy and Nefedev, Konstantin and Otsuka, Hiromi and Okabe, Yutaka},
   title = {{Large peaks in the entropy of the diluted nearest-neighbor spin-ice model on the pyrochlore lattice in a [111] magnetic field}},
   journal = {Phys. Rev. E},
   volume = {99},
   number = {2},
   pages = {022138},
   DOI = {10.1103/PhysRevE.99.022138},
   url = {https://link.aps.org/doi/10.1103/PhysRevE.99.022138},
   year = {2019},
   type = {Journal Article}
}

@article{RN124,
   author = {Assis, Michael},
   title = {{The 16-vertex model and its even and odd 8-vertex subcases on the square lattice}},
   journal = {J. Phys. A: Math. Theor.},
   volume = {50},
   number = {39},
   pages = {395001},
   ISSN = {1751-8113
1751-8121},
   DOI = {10.1088/1751-8121/aa842c},
   url = {http://dx.doi.org/10.1088/1751-8121/aa842c},
   year = {2017},
   type = {Journal Article}
}

@article{RN282,
   author = {Bañuls, Mari Carmen},
   title = {{Tensor Network Algorithms: A Route Map}},
   journal = {Annual Review of Condensed Matter Physics},
   volume = {14},
   pages = {173-191},
   ISSN = {1947-5462},
   DOI = {10.1146/annurev-conmatphys-040721-022705},
   url = {https://www.annualreviews.org/content/journals/10.1146/annurev-conmatphys-040721-022705},
   year = {2023},
   type = {Journal Article}
}

@article{RN199,
   author = {Baxter, R. J.},
   title = {{Dimers on a Rectangular Lattice}},
   journal = {J. Math. Phys.},
   volume = {9},
   number = {4},
   pages = {650-654},
   ISSN = {0022-2488},
   DOI = {10.1063/1.1664623},
   url = {https://doi.org/10.1063/1.1664623},
   year = {1968},
   type = {Journal Article}
}

@article{RN358,
   author = {Baxter, R. J.},
   title = {{Colorings of a Hexagonal Lattice}},
   journal = {J. Math. Phys.},
   volume = {11},
   number = {3},
   pages = {784-789},
   ISSN = {0022-2488},
   DOI = {10.1063/1.1665210},
   url = {https://doi.org/10.1063/1.1665210},
   year = {1970},
   type = {Journal Article}
}

@article{RN200,
   author = {Baxter, R. J.},
   title = {{Variational approximations for square lattice models in statistical mechanics}},
   journal = {J. Stat. Phys.},
   volume = {19},
   number = {5},
   pages = {461-478},
   ISSN = {1572-9613},
   DOI = {10.1007/BF01011693},
   url = {https://doi.org/10.1007/BF01011693},
   year = {1978},
   type = {Journal Article}
}

@article{RN241,
   author = {Blote, H. W. J. and Hilborst, H. J.},
   title = {{Roughening transitions and the zero-temperature triangular Ising antiferromagnet}},
   journal = {J. Phys. A: Math. Gen.},
   volume = {15},
   number = {11},
   pages = {L631},
   ISSN = {0305-4470},
   DOI = {10.1088/0305-4470/15/11/011},
   url = {https://dx.doi.org/10.1088/0305-4470/15/11/011},
   year = {1982},
   type = {Journal Article}
}

@article{RN146,
   author = {Bondy, J. A. and Welsh, D. J. A.},
   title = {{A note on the monomer dimer problem}},
   journal = {Proc. Camb. Phil. Soc.},
   volume = {62},
   number = {3},
   pages = {503-505},
   ISSN = {0305-0041},
   DOI = {10.1017/S0305004100040123},
   url = {https://doi.org/10.1017/S0305004100040123},
   year = {1966},
   type = {Journal Article}
}

@article{RN543,
   author = {Chang, T. S. and Fowler, Ralph Howard},
   title = {{Statistical theory of the adsorption of double molecules}},
   journal = {Proc. R. Soc. Lond. A},
   volume = {169},
   number = {939},
   pages = {512-531},
   DOI = {10.1098/rspa.1939.0014},
   url = {https://royalsocietypublishing.org/doi/abs/10.1098/rspa.1939.0014},
   year = {1939},
   type = {Journal Article}
}

@article{RN571,
   author = {Hwang, Chi-Ok and Kim, Seung-Yeon and Kang, Daeseung and Kim, Jin Min},
   title = {{Thermodynamic Properties of the Triangular-Lattice Ising Antiferromagnet in a Uniform Magnetic Field}},
   journal = {J. Korean Phys. Soc.},
   volume = {52},
   number = {1},
   pages = {S203-S208},
   DOI = {10.3938/jkps.52.203},
   url = {https://doi.org/10.3938/jkps.52.203},
   year = {2008},
   type = {Journal Article}
}

@phdthesis{RN345,
   author = {Colbois, Jeanne},
   title = {{Tensor network investigation of frustrated Ising models}},
   university = {EPFL},
   type = {{Ph.D. thesis (EPFL, 2022)}},
   DOI = {10.5075/epfl-thesis-9290},
   url = {https://infoscience.epfl.ch/record/293822},
   year = {2022},
   type = {Thesis}
}

@article{RN354,
   author = {Colbois, Jeanne and Hofhuis, Kevin and Luo, Zhaochu and Wang, Xueqiao and Hrabec, Aleš and Heyderman, Laura J. and Mila, Frédéric},
   title = {{Artificial out-of-plane Ising antiferromagnet on the kagome lattice with very small farther-neighbor couplings}},
   journal = {Phys. Rev. B},
   volume = {104},
   number = {2},
   pages = {024418},
   DOI = {10.1103/PhysRevB.104.024418},
   url = {https://link.aps.org/doi/10.1103/PhysRevB.104.024418},
   year = {2021},
   type = {Journal Article}
}

@article{RN65,
   author = {Fan, Chungpeng and Wu, F. Y.},
   title = {{Ising Model with Second-Neighbor Interaction. I. Some Exact Results and an Approximate Solution}},
   journal = {Phys. Rev.},
   volume = {179},
   number = {2},
   pages = {560-569},
   DOI = {10.1103/PhysRev.179.560},
   url = {https://link.aps.org/doi/10.1103/PhysRev.179.560},
   year = {1969},
   type = {Journal Article}
}

@article{RN59,
   author = {Fan, Chungpeng and Wu, F. Y.},
   title = {{General Lattice Model of Phase Transitions}},
   journal = {Phys. Rev. B},
   volume = {2},
   number = {3},
   pages = {723-733},
   DOI = {10.1103/PhysRevB.2.723},
   url = {https://link.aps.org/doi/10.1103/PhysRevB.2.723},
   year = {1970},
   type = {Journal Article}
}

@article{RN143,
   author = {Fendley, P. and Moessner, R. and Sondhi, S. L.},
   title = {{Classical dimers on the triangular lattice}},
   journal = {Phys. Rev. B},
   volume = {66},
   number = {21},
   pages = {214513},
   DOI = {10.1103/PhysRevB.66.214513},
   url = {https://link.aps.org/doi/10.1103/PhysRevB.66.214513},
   year = {2002},
   type = {Journal Article}
}

@article{RN137,
   author = {Fisher, Michael E.},
   title = {{Statistical Mechanics of Dimers on a Plane Lattice}},
   journal = {Phys. Rev.},
   volume = {124},
   number = {6},
   pages = {1664-1672},
   DOI = {10.1103/PhysRev.124.1664},
   url = {https://link.aps.org/doi/10.1103/PhysRev.124.1664},
   year = {1961},
   type = {Journal Article}
}

@article{RN471,
   author = {Fisher, Michael E.},
   title = {{Lattice Statistics‐A Review and an Exact Isotherm for a Plane Lattice Gas}},
   journal = {J. Math. Phys.},
   volume = {4},
   number = {2},
   pages = {278-286},
   ISSN = {0022-2488},
   DOI = {10.1063/1.1703952},
   url = {https://doi.org/10.1063/1.1703952},
   year = {1963},
   type = {Journal Article}
}

@article{RN349,
   author = {Fishman, M. T. and Vanderstraeten, L. and Zauner-Stauber, V. and Haegeman, J. and Verstraete, F.},
   title = {{Faster methods for contracting infinite two-dimensional tensor networks}},
   journal = {Phys. Rev. B},
   volume = {98},
   number = {23},
   pages = {235148},
   DOI = {10.1103/PhysRevB.98.235148},
   url = {https://link.aps.org/doi/10.1103/PhysRevB.98.235148},
   year = {2018},
   type = {Journal Article}
}

@article{RN360,
   author = {Fjærestad, J. O.},
   title = {{Dimer and fermionic formulations of a class of colouring problems}},
   journal = {J. Phys. A: Math. Theor.},
   volume = {45},
   number = {7},
   pages = {075001},
   ISSN = {1751-8121
1751-8113},
   DOI = {10.1088/1751-8113/45/7/075001},
   url = {https://dx.doi.org/10.1088/1751-8113/45/7/075001},
   year = {2012},
   type = {Journal Article}
}

@article{RN400,
   author = {Fowler, R. H. and Rushbrooke, G. S.},
   title = {{An attempt to extend the statistical theory of perfect solutions}},
   journal = {Trans. Faraday Soc.},
   volume = {33},
   number = {0},
   pages = {1272-1294},
   ISSN = {0014-7672},
   DOI = {10.1039/TF9373301272},
   url = {http://dx.doi.org/10.1039/TF9373301272},
   year = {1937},
   type = {Journal Article}
}

@article{RN189,
   author = {Gaunt, David S.},
   title = {{Exact Series-Expansion Study of the Monomer-Dimer Problem}},
   journal = {Phys. Rev.},
   volume = {179},
   number = {1},
   pages = {174-186},
   DOI = {10.1103/PhysRev.179.174},
   url = {https://link.aps.org/doi/10.1103/PhysRev.179.174},
   year = {1969},
   type = {Journal Article}
}

@article{RN356,
   author = {Giuliani, Alessandro and Jauslin, Ian and Lieb, Elliott H.},
   title = {{A Pfaffian Formula for Monomer–Dimer Partition Functions}},
   journal = {J. Stat. Phys.},
   volume = {163},
   number = {2},
   pages = {211-238},
   ISSN = {1572-9613},
   DOI = {10.1007/s10955-016-1484-1},
   url = {https://doi.org/10.1007/s10955-016-1484-1},
   year = {2016},
   type = {Journal Article}
}

@article{RN281,
   author = {Haegeman, Jutho and Verstraete, Frank},
   title = {{Diagonalizing Transfer Matrices and Matrix Product Operators: A Medley of Exact and Computational Methods}},
   journal = {Annual Review of Condensed Matter Physics},
   volume = {8},
   pages = {355-406},
   ISSN = {1947-5462},
   DOI = {10.1146/annurev-conmatphys-031016-025507},
   url = {https://www.annualreviews.org/content/journals/10.1146/annurev-conmatphys-031016-025507},
   year = {2017},
   type = {Journal Article}
}

@article{RN148,
   author = {Heilmann, Ole J. and Lieb, Elliott H.},
   title = {{Monomers and Dimers}},
   journal = {Phys. Rev. Lett.},
   volume = {24},
   number = {25},
   pages = {1412-1414},
   DOI = {10.1103/PhysRevLett.24.1412},
   url = {https://link.aps.org/doi/10.1103/PhysRevLett.24.1412},
   year = {1970},
   type = {Journal Article}
}

@article{RN149,
   author = {Heilmann, Ole J. and Lieb, Elliott H.},
   title = {{Theory of monomer-dimer systems}},
   journal = {Commun. Math. Phys.},
   volume = {25},
   number = {3},
   pages = {190-232},
   ISSN = {1432-0916},
   DOI = {10.1007/BF01877590},
   url = {https://doi.org/10.1007/BF01877590},
   year = {1972},
   type = {Journal Article}
}

@article{RN568,
   author = {Hwang, Chi-Ok and Kim, Seung-Yeon},
   title = {{Yang–Lee zeros of triangular Ising antiferromagnets}},
   journal = {Physica A},
   volume = {389},
   number = {24},
   pages = {5650-5654},
   ISSN = {0378-4371},
   DOI = {10.1016/j.physa.2010.08.050},
   url = {https://www.sciencedirect.com/science/article/pii/S0378437110007533},
   year = {2010},
   type = {Journal Article}
}

@article{RN104,
   author = {Isakov, S. V. and Raman, K. S. and Moessner, R. and Sondhi, S. L.},
   title = {{Magnetization curve of spin ice in a [111] magnetic field}},
   journal = {Phys. Rev. B},
   volume = {70},
   number = {10},
   pages = {104418},
   DOI = {10.1103/PhysRevB.70.104418},
   url = {https://link.aps.org/doi/10.1103/PhysRevB.70.104418},
   year = {2004},
   type = {Journal Article}
}

@article{RN387,
   author = {Izmailian, N. Sh and Hu, Chin-Kun and Kenna, R.},
   title = {{Exact solution of the dimer model on the generalized finite checkerboard lattice}},
   journal = {Phys. Rev. E},
   volume = {91},
   number = {6},
   pages = {062139},
   DOI = {10.1103/PhysRevE.91.062139},
   url = {https://link.aps.org/doi/10.1103/PhysRevE.91.062139},
   year = {2015},
   type = {Journal Article}
}

@article{RN355,
   author = {Jerrum, Mark},
   title = {{Two-dimensional monomer-dimer systems are computationally intractable}},
   journal = {J. Stat. Phys.},
   volume = {48},
   number = {1},
   pages = {121-134},
   ISSN = {1572-9613},
   DOI = {10.1007/BF01010403},
   url = {https://doi.org/10.1007/BF01010403},
   year = {1987},
   type = {Journal Article}
}

@article{RN82,
   author = {Kanô, Kenzi and Naya, Shigeo},
   title = {{Antiferromagnetism. The Kagomé Ising Net}},
   journal = {Prog. Theor. Phys.},
   volume = {10},
   number = {2},
   pages = {158-172},
   ISSN = {0033-068X},
   DOI = {10.1143/ptp/10.2.158},
   url = {https://doi.org/10.1143/ptp/10.2.158},
   year = {1953},
   type = {Journal Article}
}

@article{RN136,
   author = {Kasteleyn, P. W.},
   title = {{The statistics of dimers on a lattice: I. The number of dimer arrangements on a quadratic lattice}},
   journal = {Physica},
   volume = {27},
   number = {12},
   pages = {1209-1225},
   ISSN = {0031-8914},
   DOI = {10.1016/0031-8914(61)90063-5},
   url = {https://www.sciencedirect.com/science/article/pii/0031891461900635},
   year = {1961},
   type = {Journal Article}
}

@article{RN139,
   author = {Kasteleyn, P. W.},
   title = {{Dimer Statistics and Phase Transitions}},
   journal = {J. Math. Phys.},
   volume = {4},
   number = {2},
   pages = {287-293},
   ISSN = {0022-2488},
   DOI = {10.1063/1.1703953},
   url = {https://doi.org/10.1063/1.1703953},
   year = {1963},
   type = {Journal Article}
}

@article{RN533,
   author = {Kattemölle, Joris},
   title = {{Edge coloring lattice graphs}},
   journal = {J. Math. Phys.},
   volume = {66},
   number = {5},
   pages = {051901},
   ISSN = {0022-2488},
   DOI = {10.1063/5.0243007},
   url = {https://doi.org/10.1063/5.0243007},
   year = {2025},
   type = {Journal Article}
}

@article{RN569,
   author = {Kim, Seung-Yeon},
   title = {{Ising antiferromagnets on honeycomb and square lattices in the critical magnetic field}},
   journal = {J. Korean Phys. Soc.},
   volume = {61},
   number = {12},
   pages = {1950-1955},
   ISSN = {1976-8524},
   DOI = {10.3938/jkps.61.1950},
   url = {https://doi.org/10.3938/jkps.61.1950},
   year = {2012},
   type = {Journal Article}
}

@article{RN579,
   author = {Kong, Yong},
   title = {{Recurrence solution of monomer-polymer models on two-dimensional rectangular lattices}},
   journal = {Phys. Rev. E},
   volume = {110},
   number = {5},
   pages = {054135},
   DOI = {10.1103/PhysRevE.110.054135},
   url = {https://link.aps.org/doi/10.1103/PhysRevE.110.054135},
   year = {2024},
   type = {Journal Article}
}

@article{RN570,
   author = {Lebowitz, Joel L. and Ruelle, David and Speer, Eugene R.},
   title = {{Location of the Lee-Yang zeros and absence of phase transitions in some Ising spin systems}},
   journal = {J. Math. Phys.},
   volume = {53},
   number = {9},
   pages = {095211},
   ISSN = {0022-2488},
   DOI = {10.1063/1.4738622},
   url = {https://doi.org/10.1063/1.4738622},
   year = {2012},
   type = {Journal Article}
}

@article{RN57,
   author = {Lee, T. D. and Yang, C. N.},
   title = {{Statistical Theory of Equations of State and Phase Transitions. II. Lattice Gas and Ising Model}},
   journal = {Phys. Rev.},
   volume = {87},
   number = {3},
   pages = {410-419},
   DOI = {10.1103/PhysRev.87.410},
   url = {https://link.aps.org/doi/10.1103/PhysRev.87.410},
   year = {1952},
   type = {Journal Article}
}

@article{RN225,
   author = {Li, De-Zhang and Huang, Wei-Jie and Yao, Yao and Yang, Xiao-Bao},
   title = {{Exact results for the residual entropy of ice hexagonal monolayer}},
   journal = {Phys. Rev. E},
   volume = {107},
   number = {5},
   pages = {054121},
   DOI = {10.1103/PhysRevE.107.054121},
   url = {https://link.aps.org/doi/10.1103/PhysRevE.107.054121},
   year = {2023},
   type = {Journal Article}
}

@article{RN558,
   author = {Li, De-Zhang and Wang, Xin and Yang, Xiao-Bao},
   title = {{Free-Fermion Models and Two-Dimensional Ising Models Under Zero Field and Imaginary Field i($\pi$/2)kBT}},
   journal = {Entropy},
   volume = {27},
   number = {8},
   pages = {799},
   ISSN = {1099-4300},
   DOI = {10.3390/e27080799},
   url = {https://www.mdpi.com/1099-4300/27/8/799},
   year = {2025},
   type = {Journal Article}
}

@article{RN141,
   author = {Lieb, Elliott H.},
   title = {{Solution of the Dimer Problem by the Transfer Matrix Method}},
   journal = {J. Math. Phys.},
   volume = {8},
   number = {12},
   pages = {2339-2341},
   ISSN = {0022-2488},
   DOI = {10.1063/1.1705163},
   url = {https://doi.org/10.1063/1.1705163},
   year = {1967},
   type = {Journal Article}
}

@article{RN380,
   author = {Loh, Y. L. and Yao, Dao-Xin and Carlson, E. W.},
   title = {{Dimers on the triangular kagome lattice}},
   journal = {Phys. Rev. B},
   volume = {78},
   number = {22},
   pages = {224410},
   DOI = {10.1103/PhysRevB.78.224410},
   url = {https://link.aps.org/doi/10.1103/PhysRevB.78.224410},
   year = {2008},
   type = {Journal Article}
}

@article{RN486,
   author = {Lu, Wentao T. and Wu, F. Y.},
   title = {{Close-packed dimers on nonorientable surfaces}},
   journal = {Phys. Lett. A},
   volume = {293},
   number = {5},
   pages = {235-246},
   ISSN = {0375-9601},
   DOI = {10.1016/S0375-9601(02)00019-1},
   url = {https://www.sciencedirect.com/science/article/pii/S0375960102000191},
   year = {2002},
   type = {Journal Article}
}

@book{RN465,
   author = {McCoy, B.M. and Wu, T.T.},
   title = {{The Two-Dimensional Ising Model: Second Edition}},
   publisher = {Dover Publications},
   address = {New York},
   ISBN = {9780486493350},
   year = {2014},
   type = {Book}
}

@article{RN399,
   author = {Metcalf, B. D. and Yang, C. P.},
   title = {{Degeneracy of antiferromagnetic Ising lattices at critical magnetic field and zero temperature}},
   journal = {Phys. Rev. B},
   volume = {18},
   number = {5},
   pages = {2304-2307},
   DOI = {10.1103/PhysRevB.18.2304},
   url = {https://link.aps.org/doi/10.1103/PhysRevB.18.2304},
   year = {1978},
   type = {Journal Article}
}

@article{RN544,
   author = {Miller, A. R.},
   title = {{The number of configurations of a cooperative assembly}},
   journal = {Proc. Camb. Phil. Soc.},
   volume = {38},
   number = {1},
   pages = {109-124},
   ISSN = {0305-0041},
   DOI = {10.1017/S030500410002226X},
   url = {https://www.cambridge.org/core/product/29AF14D933FBB99958E7ADF62B6554A5},
   year = {1942},
   type = {Journal Article}
}

@article{RN157,
   author = {Moessner, R. and Sondhi, S. L.},
   title = {{Ising models of quantum frustration}},
   journal = {Phys. Rev. B},
   volume = {63},
   number = {22},
   pages = {224401},
   DOI = {10.1103/PhysRevB.63.224401},
   url = {https://link.aps.org/doi/10.1103/PhysRevB.63.224401},
   year = {2001},
   type = {Journal Article}
}

@article{RN158,
   author = {Moessner, R. and Sondhi, S. L.},
   title = {{Theory of the [111] magnetization plateau in spin ice}},
   journal = {Phys. Rev. B},
   volume = {68},
   number = {6},
   pages = {064411},
   DOI = {10.1103/PhysRevB.68.064411},
   url = {https://link.aps.org/doi/10.1103/PhysRevB.68.064411},
   year = {2003},
   type = {Journal Article}
}

@article{RN385,
   author = {Moessner, R. and Sondhi, S. L. and Fradkin, Eduardo},
   title = {{Short-ranged resonating valence bond physics, quantum dimer models, and Ising gauge theories}},
   journal = {Phys. Rev. B},
   volume = {65},
   number = {2},
   pages = {024504},
   DOI = {10.1103/PhysRevB.65.024504},
   url = {https://link.aps.org/doi/10.1103/PhysRevB.65.024504},
   year = {2001},
   type = {Journal Article}
}

@inbook{RN397,
   author = {Montroll, Elliott W.},
   title = {{Lattice Statistics}},
   booktitle = {Applied Combinatorial Mathematics},
   publisher = {Wiley},
   address = {New York},
   year = {1964},
   type = {Book Section}
}

@article{RN357,
   author = {Morita, Satoshi and Lee, Hyun-Yong and Damle, Kedar and Kawashima, Naoki},
   title = {{Ashkin-Teller phase transition and multicritical behavior in a classical monomer-dimer model}},
   journal = {Phys. Rev. Research},
   volume = {5},
   number = {4},
   pages = {043061},
   DOI = {10.1103/PhysRevResearch.5.043061},
   url = {https://link.aps.org/doi/10.1103/PhysRevResearch.5.043061},
   year = {2023},
   type = {Journal Article}
}

@article{RN147,
   author = {Nagle, John F.},
   title = {{New Series-Expansion Method for the Dimer Problem}},
   journal = {Phys. Rev.},
   volume = {152},
   number = {1},
   pages = {190-197},
   DOI = {10.1103/PhysRev.152.190},
   url = {https://link.aps.org/doi/10.1103/PhysRev.152.190},
   year = {1966},
   type = {Journal Article}
}

@article{RN203,
   author = {Okunishi, Kouichi and Nishino, Tomotoshi and Ueda, Hiroshi},
   title = {{Developments in the Tensor Network — from Statistical Mechanics to Quantum Entanglement}},
   journal = {J. Phys. Soc. Jpn.},
   volume = {91},
   number = {6},
   pages = {062001},
   ISSN = {0031-9015},
   DOI = {10.7566/JPSJ.91.062001},
   url = {https://doi.org/10.7566/JPSJ.91.062001},
   year = {2022},
   type = {Journal Article}
}

@article{RN545,
   author = {Orr, W. J. C.},
   title = {{On the calculation of certain higher-order Bethe approximations}},
   journal = {Trans. Faraday Soc.},
   volume = {40},
   number = {0},
   pages = {306-320},
   ISSN = {0014-7672},
   DOI = {10.1039/TF9444000306},
   url = {http://dx.doi.org/10.1039/TF9444000306},
   year = {1944},
   type = {Journal Article}
}

@article{RN151,
   author = {Otsuka, Hiromi},
   title = {{Monomer-Dimer Mixture on a Honeycomb Lattice}},
   journal = {Phys. Rev. Lett.},
   volume = {106},
   number = {22},
   pages = {227204},
   DOI = {10.1103/PhysRevLett.106.227204},
   url = {https://link.aps.org/doi/10.1103/PhysRevLett.106.227204},
   year = {2011},
   type = {Journal Article}
}

@phdthesis{RN347,
   author = {Otsuka, Takahiro},
   title = {{The monomer-dimer models in two and three dimensions: Tensor renormalization group study}},
   university = {Osaka University},
   type = {{Ph.D. thesis (Osaka University, 2022)}},
   DOI = {10.18910/87816},
   year = {2022},
   type = {Thesis}
}

@article{RN393,
   author = {Pearce, Paul A. and Vittorini-Orgeas, Alessandra},
   title = {{Yang–Baxter solution of dimers as a free-fermion six-vertex model}},
   journal = {J. Phys. A: Math. Theor.},
   volume = {50},
   number = {43},
   pages = {434001},
   ISSN = {1751-8121
1751-8113},
   DOI = {10.1088/1751-8121/aa86bc},
   url = {https://dx.doi.org/10.1088/1751-8121/aa86bc},
   year = {2017},
   type = {Journal Article}
}

@article{RN407,
   author = {Phares, Alain J.},
   title = {{The occupation statistics for indistinguishable dumbbells on a rectangular lattice space. I}},
   journal = {J. Math. Phys.},
   volume = {25},
   number = {6},
   pages = {1756-1770},
   ISSN = {0022-2488},
   DOI = {10.1063/1.526350},
   url = {https://doi.org/10.1063/1.526350},
   year = {1984},
   type = {Journal Article}
}

@article{RN408,
   author = {Phares, Alain J. and Shaw, Donald E. and Wunderlich, Francis J.},
   title = {{An approximate solution of the monomer–dimer problem on a square lattice. II}},
   journal = {J. Math. Phys.},
   volume = {26},
   number = {7},
   pages = {1762-1768},
   ISSN = {0022-2488},
   DOI = {10.1063/1.526888},
   url = {https://doi.org/10.1063/1.526888},
   year = {1985},
   type = {Journal Article}
}

@article{RN577,
   author = {Roberts, John Keith},
   title = {{Composite films of oxygen and hydrogen on tungsten}},
   journal = {Proc. R. Soc. Lond. A},
   volume = {152},
   number = {876},
   pages = {477-480},
   DOI = {10.1098/rspa.1935.0202},
   url = {https://royalsocietypublishing.org/doi/abs/10.1098/rspa.1935.0202},
   year = {1935},
   type = {Journal Article}
}

@article{RN578,
   author = {Roberts, J. K. and Miller, A. R.},
   title = {{The application of statistical methods to immobile adsorbed films}},
   journal = {Proc. Camb. Phil. Soc.},
   volume = {35},
   number = {2},
   pages = {293-297},
   ISSN = {0305-0041},
   DOI = {10.1017/S0305004100020971},
   url = {https://www.cambridge.org/core/product/159C79FB1958622125A97E39BD48B624},
   year = {1939},
   type = {Journal Article}
}

@article{RN375,
   author = {Samuel, Stuart},
   title = {{The use of anticommuting variable integrals in statistical mechanics. III. Unsolved models}},
   journal = {J. Math. Phys.},
   volume = {21},
   number = {12},
   pages = {2820-2833},
   ISSN = {0022-2488},
   DOI = {10.1063/1.524406},
   url = {https://doi.org/10.1063/1.524406},
   year = {1980},
   type = {Journal Article}
}

@article{RN567,
   author = {Semjan, M and Žukovič, M},
   title = {{Magnetocaloric Properties of an Ising Antiferromagnet on a Kagome Lattice}},
   journal = {Acta Phys. Pol. A},
   volume = {137},
   number = {5},
   pages = {622-624},
   ISSN = {0587-4246},
   DOI = {10.12693/APhysPolA.137.622},
   url = {http://przyrbwn.icm.edu.pl/APP/ABSTR/137/a137-5-12.html},
   year = {2020},
   type = {Journal Article}
}

@article{RN336,
   author = {Song, Feng-Feng and Lin, Tong-Yu and Zhang, Guang-Ming},
   title = {{General tensor network theory for frustrated classical spin models in two dimensions}},
   journal = {Phys. Rev. B},
   volume = {108},
   number = {22},
   pages = {224404},
   DOI = {10.1103/PhysRevB.108.224404},
   url = {https://link.aps.org/doi/10.1103/PhysRevB.108.224404},
   year = {2023},
   type = {Journal Article}
}

@article{RN121,
   author = {Syôzi, Itiro},
   title = {{Statistics of Kagomé Lattice}},
   journal = {Prog. Theor. Phys.},
   volume = {6},
   number = {3},
   pages = {306-308},
   ISSN = {0033-068X},
   DOI = {10.1143/ptp/6.3.306},
   url = {https://doi.org/10.1143/ptp/6.3.306},
   year = {1951},
   type = {Journal Article}
}

@article{RN138,
   author = {Temperley, H. N. V. and Fisher, Michael E.},
   title = {{Dimer problem in statistical mechanics-an exact result}},
   journal = {Phil. Mag.},
   volume = {6},
   number = {68},
   pages = {1061-1063},
   ISSN = {0031-8086},
   DOI = {10.1080/14786436108243366},
   url = {https://doi.org/10.1080/14786436108243366},
   year = {1961},
   type = {Journal Article}
}

@article{RN401,
   author = {Tzeng, W. J. and Wu, F. Y.},
   title = {{Dimers on a Simple-Quartic Net with a Vacancy}},
   journal = {J. Stat. Phys.},
   volume = {110},
   number = {3},
   pages = {671-689},
   ISSN = {1572-9613},
   DOI = {10.1023/A:1022155701655},
   url = {https://doi.org/10.1023/A:1022155701655},
   year = {2003},
   type = {Journal Article}
}

@article{RN352,
   author = {Vanderstraeten, Laurens and Haegeman, Jutho and Verstraete, Frank},
   title = {{Tangent-space methods for uniform matrix product states}},
   journal = {SciPost Phys. Lect. Notes},
   volume = {7},
   pages = {1-77},
   ISSN = {2590-1990},
   DOI = {10.21468/SciPostPhysLectNotes.7},
   url = {https://doi.org/10.21468/SciPostPhysLectNotes.7},
   year = {2019},
   type = {Journal Article}
}

@article{RN103,
   author = {Vanderstraeten, Laurens and Vanhecke, Bram and Verstraete, Frank},
   title = {{Residual entropies for three-dimensional frustrated spin systems with tensor networks}},
   journal = {Phys. Rev. E},
   volume = {98},
   number = {4},
   pages = {042145},
   DOI = {10.1103/PhysRevE.98.042145},
   url = {https://link.aps.org/doi/10.1103/PhysRevE.98.042145},
   year = {2018},
   type = {Journal Article}
}

@article{RN181,
   author = {Vanhecke, Bram and Colbois, Jeanne and Vanderstraeten, Laurens and Verstraete, Frank and Mila, Frédéric},
   title = {{Solving frustrated Ising models using tensor networks}},
   journal = {Phys. Rev. Research},
   volume = {3},
   number = {1},
   pages = {013041},
   DOI = {10.1103/PhysRevResearch.3.013041},
   url = {https://link.aps.org/doi/10.1103/PhysRevResearch.3.013041},
   year = {2021},
   type = {Journal Article}
}

@article{RN144,
   author = {Wang, Fa and Wu, F. Y.},
   title = {{Exact solution of close-packed dimers on the kagomé lattice}},
   journal = {Phys. Rev. E},
   volume = {75},
   number = {4},
   pages = {040105},
   DOI = {10.1103/PhysRevE.75.040105},
   url = {https://link.aps.org/doi/10.1103/PhysRevE.75.040105},
   year = {2007},
   type = {Journal Article}
}

@article{RN81,
   author = {Wannier, G. H.},
   title = {{Antiferromagnetism. The Triangular Ising Net}},
   journal = {Phys. Rev.},
   volume = {79},
   number = {2},
   pages = {357-364},
   DOI = {10.1103/PhysRev.79.357},
   url = {https://link.aps.org/doi/10.1103/PhysRev.79.357},
   year = {1950},
   type = {Journal Article}
}

@article{RN341,
   author = {White, Steven R.},
   title = {{Density matrix formulation for quantum renormalization groups}},
   journal = {Phys. Rev. Lett.},
   volume = {69},
   number = {19},
   pages = {2863-2866},
   DOI = {10.1103/PhysRevLett.69.2863},
   url = {https://link.aps.org/doi/10.1103/PhysRevLett.69.2863},
   year = {1992},
   type = {Journal Article}
}

@article{RN389,
   author = {Wildeboer, Julia and Nussinov, Zohar and Seidel, Alexander},
   title = {{Exact solution and correlations of a dimer model on the checkerboard lattice}},
   journal = {Phys. Rev. B},
   volume = {102},
   number = {2},
   pages = {020401},
   DOI = {10.1103/PhysRevB.102.020401},
   url = {https://link.aps.org/doi/10.1103/PhysRevB.102.020401},
   year = {2020},
   type = {Journal Article}
}

@article{RN150,
   author = {Wood, D. W. and Goldfinch, M.},
   title = {{Vertex models for the hard-square and hard-hexagon gases, and critical parameters from the scaling transformation}},
   journal = {J. Phys. A: Math. Gen.},
   volume = {13},
   number = {8},
   pages = {2781-2794},
   ISSN = {0305-4470
1361-6447},
   DOI = {10.1088/0305-4470/13/8/026},
   url = {http://dx.doi.org/10.1088/0305-4470/13/8/026},
   year = {1980},
   type = {Journal Article}
}

@article{RN140,
   author = {Wu, F. Y.},
   title = {{Remarks on the Modified Potassium Dihydrogen Phosphate Model of a Ferroelectric}},
   journal = {Phys. Rev.},
   volume = {168},
   number = {2},
   pages = {539-543},
   DOI = {10.1103/PhysRev.168.539},
   url = {https://link.aps.org/doi/10.1103/PhysRev.168.539},
   year = {1968},
   type = {Journal Article}
}

@article{RN117,
   author = {Wu, F. Y.},
   title = {{Eight-vertex model on the honeycomb lattice}},
   journal = {J. Math. Phys.},
   volume = {15},
   number = {6},
   pages = {687-691},
   ISSN = {0022-2488},
   DOI = {10.1063/1.1666712},
   url = {https://doi.org/10.1063/1.1666712},
   year = {1974},
   type = {Journal Article}
}

@article{RN212,
   author = {Wu, F. Y.},
   title = {{The Potts model}},
   journal = {Rev. Mod. Phys.},
   volume = {54},
   number = {1},
   pages = {235-268},
   DOI = {10.1103/RevModPhys.54.235},
   url = {https://link.aps.org/doi/10.1103/RevModPhys.54.235},
   year = {1982},
   type = {Journal Article}
}

@article{RN145,
   author = {Wu, F. Y.},
   title = {{DIMERS ON TWO-DIMENSIONAL LATTICES}},
   journal = {Int. J. Mod. Phys. B},
   volume = {20},
   number = {32},
   pages = {5357-5371},
   ISSN = {0217-9792},
   DOI = {10.1142/S0217979206036478},
   url = {https://doi.org/10.1142/S0217979206036478},
   year = {2006},
   type = {Journal Article}
}

@article{RN402,
   author = {Wu, F. Y.},
   title = {{Pfaffian solution of a dimer-monomer problem: Single monomer on the boundary}},
   journal = {Phys. Rev. E},
   volume = {74},
   number = {2},
   pages = {020104},
   DOI = {10.1103/PhysRevE.74.020104},
   url = {https://link.aps.org/doi/10.1103/PhysRevE.74.020104},
   year = {2006},
   type = {Journal Article}
}

@article{RN129,
   author = {Wu, F. Y. and Kunz, H.},
   title = {{The Odd Eight-Vertex Model}},
   journal = {J. Stat. Phys.},
   volume = {116},
   number = {1},
   pages = {67-78},
   ISSN = {1572-9613},
   DOI = {10.1023/B:JOSS.0000037206.47155.58},
   url = {https://doi.org/10.1023/B:JOSS.0000037206.47155.58},
   year = {2004},
   type = {Journal Article}
}

@article{RN476,
   author = {Wu, F. Y. and Tzeng, Wen-Jer and Izmailian, N. Sh},
   title = {{Exact solution of a monomer-dimer problem: A single boundary monomer on a nonbipartite lattice}},
   journal = {Phys. Rev. E},
   volume = {83},
   number = {1},
   pages = {011106},
   DOI = {10.1103/PhysRevE.83.011106},
   url = {https://link.aps.org/doi/10.1103/PhysRevE.83.011106},
   year = {2011},
   type = {Journal Article}
}

@article{RN382,
   author = {Wu, F. Y. and Wang, Fa},
   title = {{Dimers on the kagome lattice I: Finite lattices}},
   journal = {Physica A},
   volume = {387},
   number = {16},
   pages = {4148-4156},
   ISSN = {0378-4371},
   DOI = {10.1016/j.physa.2008.02.054},
   url = {https://www.sciencedirect.com/science/article/pii/S0378437108002021},
   year = {2008},
   type = {Journal Article}
}

@article{RN351,
   author = {Zauner-Stauber, V. and Vanderstraeten, L. and Fishman, M. T. and Verstraete, F. and Haegeman, J.},
   title = {{Variational optimization algorithms for uniform matrix product states}},
   journal = {Phys. Rev. B},
   volume = {97},
   number = {4},
   pages = {045145},
   DOI = {10.1103/PhysRevB.97.045145},
   url = {https://link.aps.org/doi/10.1103/PhysRevB.97.045145},
   year = {2018},
   type = {Journal Article}
}

@article{RN576,
   author = {Zhao, H. H. and Xie, Z. Y. and Chen, Q. N. and Wei, Z. C. and Cai, J. W. and Xiang, T.},
   title = {{Renormalization of tensor-network states}},
   journal = {Phys. Rev. B},
   volume = {81},
   number = {17},
   pages = {174411},
   DOI = {10.1103/PhysRevB.81.174411},
   url = {https://link.aps.org/doi/10.1103/PhysRevB.81.174411},
   year = {2010},
   type = {Journal Article}
}

@article{RN581,
   author = {Temperley, H. N. V.},
   title = {{Application of the Mayer Method to the Melting Problem}},
   journal = {Proc. Phys. Soc.},
   volume = {74},
   number = {2},
   pages = {183-195},
   ISSN = {0370-1328},
   DOI = {10.1088/0370-1328/74/2/306},
   url = {https://dx.doi.org/10.1088/0370-1328/74/2/306},
   year = {1959},
   type = {Journal Article}
}

@article{RN582,
   author = {Temperley, H. N. V.},
   title = {{On the Asymptotic Behaviour of the Mayer Cluster Series in the Antiferromagnetic Problem}},
   journal = {Proc. Phys. Soc.},
   volume = {74},
   number = {4},
   pages = {432-443},
   ISSN = {0370-1328},
   DOI = {10.1088/0370-1328/74/4/307},
   url = {https://dx.doi.org/10.1088/0370-1328/74/4/307},
   year = {1959},
   type = {Journal Article}
}

@article{RN594,
   author = {Nietner, Alexander and Vanhecke, Bram and Verstraete, Frank and Eisert, Jens and Vanderstraeten, Laurens},
   title = {{Efficient variational contraction of two-dimensional tensor networks with a non-trivial unit cell}},
   journal = {Quantum},
   volume = {4},
   pages = {328},
   DOI = {10.22331/q-2020-09-21-328},
   url = {https://doi.org/10.22331/q-2020-09-21-328},
   year = {2020},
   type = {Journal Article}
}

@article{RN660,
   author = {Chung, Sanghyeok and Kim, Hyoungjun and Lee, Seungeun and Lee, Suinne and Oh, Seungsang},
   title = {{Honeycomb-lattice monomer-dimer mixtures}},
   journal = {J. Stat. Phys.},
   volume = {192},
   number = {11},
   pages = {149},
   ISSN = {1572-9613},
   DOI = {10.1007/s10955-025-03535-5},
   url = {https://doi.org/10.1007/s10955-025-03535-5},
   year = {2025},
   type = {Journal Article}
}

@article{RN671,
   author = {Li, De-Zhang and Wang, Xin},
   title = {{Free-fermion approach to the partition function zeros: Special boundary conditions and product form of solution}},
   journal = {Phys. Rev. Research},
   volume = {7},
   number = {4},
   pages = {043258},
   DOI = {10.1103/b6d1-6sk5},
   url = {https://link.aps.org/doi/10.1103/b6d1-6sk5},
   year = {2025},
   type = {Journal Article}
}

@article{RN733,
   author = {Xu, Xia-Ze and Lin, Tong-Yu and Zhang, Guang-Ming},
   title = {{Equivalence of residual entropy of hexagonal and cubic ices from tensor network methods}},
   journal = {Phys. Rev. B},
   volume = {113},
   number = {21},
   pages = {214416},
   DOI = {10.1103/3myh-6pdf},
   url = {https://link.aps.org/doi/10.1103/3myh-6pdf},
   year = {2026},
   type = {Journal Article}
}

@article{RN694,
   author = {Xie, Z. Y. and Chen, J. and Qin, M. P. and Zhu, J. W. and Yang, L. P. and Xiang, T.},
   title = {{Coarse-graining renormalization by higher-order singular value decomposition}},
   journal = {Phys. Rev. B},
   volume = {86},
   number = {4},
   pages = {045139},
   DOI = {10.1103/PhysRevB.86.045139},
   url = {https://link.aps.org/doi/10.1103/PhysRevB.86.045139},
   year = {2012},
   type = {Journal Article}
}

@article{RN768,
   author = {White, Steven R.},
   title = {{Density-matrix algorithms for quantum renormalization groups}},
   journal = {Phys. Rev. B},
   volume = {48},
   number = {14},
   pages = {10345-10356},
   DOI = {10.1103/PhysRevB.48.10345},
   url = {https://link.aps.org/doi/10.1103/PhysRevB.48.10345},
   year = {1993},
   type = {Journal Article}
}

\end{document}